\documentclass[authoryear,preprint,12pt]{elsarticle}

\usepackage{graphicx}
\usepackage{booktabs}
\usepackage{threeparttable}
\graphicspath{{figures/}}
\usepackage{amssymb}
\usepackage{amsmath}
\usepackage{algorithm}
\usepackage{algorithmic}
\usepackage{xurl}
\usepackage[hidelinks]{hyperref}

\journal{Engineering Applications of Artificial Intelligence}

\begin{document}

\begin{frontmatter}



\title{Sharing the Control Authority Between Deep Reinforcement Learning and Model Predictive Control: Application to Multi-Class Transportation Networks}


\author{Giray Onur, Azita Dabiri, Bart De Schutter} 

\affiliation{organization={Delft Center for Systems and Control, Delft University of Technology},
            city={Delft},
            postcode={2628CD}, 
            country={The Netherlands}}

\begin{abstract}
Transportation networks, in particular multi-class transportation networks (i.e., networks with mixed vehicle types), are complex systems that are challenging to control. Recently, Deep Reinforcement Learning (DRL), which learns control policies from interactions with the environment, and Model Predictive Control (MPC), which uses a system model to optimize control inputs, have been increasingly utilized for transportation network control. However, nonlinear system dynamics and high-dimensional state spaces in large-scale networks limit DRL’s learning capacity under time-constrained training and increase MPC’s computation time, hindering real-time implementation with limited computational resources. Moreover, MPC depends on an accurate network model, which is often unavailable for complex systems such as multi-class transportation networks. This paper proposes a novel DRL-MPC framework for multi-class transportation networks that divides control authority between DRL and MPC, combining DRL’s fast online computation and model independence with MPC’s built-in optimization and constraint-handling capabilities. In the hierarchical framework, MPC operates at the higher level and determines low-frequency control inputs whose slower update rate accommodates its high computation time, while DRL operates at the lower level and determines high-frequency control inputs using its fast online deployment. The framework is evaluated on a multi-class freeway network against a hierarchical MPC controller and a hybrid state-feedback-MPC controller, including scenarios with model mismatch and noisy traffic demands. Results show that the proposed framework outperforms the hybrid state-feedback-MPC controller, substantially reduces online computation time compared with the hierarchical MPC controller, and provides more effective constraint enforcement under model mismatch.
\end{abstract}




\end{frontmatter}



\section{Introduction}
\label{sec:intro}

Rapid and often unregulated growth in modern cities has led to significant challenges, one of the most critical being traffic congestion, which negatively impacts mobility and leads to various environmental, social, and economic issues, such as increased pollution, fuel consumption, noise, accidents, and longer travel times. These challenges have spurred a surge in the use of advanced control methods, such as Model Predictive Control (MPC) and Deep Reinforcement Learning (DRL), to improve the efficiency of transportation networks.

MPC has been widely applied to transportation networks (e.g., controlling ramp metering rates, variable speed limits, and route guidance) due to its ability to handle constraints and to optimize long-term objectives \citep{Mayne2000,Rawlings2018} (e.g., reducing the total time spent (TTS) or the fuel consumption). At each control step, MPC predicts the future evolution of the system using a prediction model over a finite horizon, solves an optimization problem to compute the optimal control input trajectory, applies only the first control input of the computed trajectory, and repeats this procedure at the next control step in a receding-horizon fashion. MPC approaches have been applied to the control of single-class \citep{todorovic2020distributed,van2018efficient} and multi-class freeway networks \citep{liu2016model,pasquale2017multi}, as well as to urban transportation networks \citep{pham2023distributed,sirmatel2017economic}. However, the performance of MPC strongly depends on the accuracy of the network model and can degrade significantly in the presence of model mismatch. In addition, the computational complexity of solving MPC problems in real time with limited computational resources poses a major challenge, particularly when dealing with nonlinear, high-dimensional systems such as large-scale multi-class transportation networks \citep{Castaneda2022,Siri2021}. Moreover, since the MPC controller then relies on a nonlinear optimizer to compute the control inputs, the resulting solutions are often suboptimal due to the nonconvex nature of the optimization problem, which is often mitigated by using a multi-start approach \citep{hegyi2005model}, resulting however in longer computation time.

DRL, on the other hand, has emerged as an effective model-free approach that can learn control policies directly from interactions with the environment \citep{Arulkumaran2017,Mnih2013}, making it particularly appealing in scenarios with limited model knowledge or uncertain system dynamics, as it does not require an explicit system model. This has led to the application of DRL in various transportation network management tasks, including ramp metering control \citep{belletti2017expert,han2022physics} and variable speed limit control \citep{gao2024variable,kuvsic2020overview} in freeway networks, along with traffic signal control in urban networks \citep{liang2019deep,wu2020multi}. Despite its potential, DRL faces several challenges, including poor sample efficiency and difficulty in handling hard constraints \citep{DulacArnold2021}. Poor sample efficiency indicates the need for many interactions with the environment to learn a good policy, which can make policy learning in transportation networks time consuming, as these interactions must often be obtained through computationally intensive simulations or costly field data collection. Furthermore, hard operational constraints such as vehicle queue length limits are not inherently ensured for standard DRL frameworks \citep{Haydari2022}.

The complementary strengths and limitations of MPC and DRL have motivated the development of hybrid control frameworks that integrate both approaches. In existing hybrid approaches, MPC is often employed as a safety filter to ensure constraint satisfaction while preserving the model-free nature of DRL \citep{zanon2020safe}, or DRL is used to adjust MPC’s control inputs or parameters to compensate for inaccuracies in the prediction model \citep{Gros2020}. Similar strategies have also been applied to mitigate model errors in ramp metering control within freeway networks \citep{airaldi2025reinforcement,sun2024novel}. However, despite overall progress, integrating DRL and MPC remains a developing area with a limited number of practical applications.


In transportation network control, updating different control inputs at different frequencies has been a common strategy, since it limits the computational burden of recomputing every input at the highest frequency.
For example, in the framework proposed by \citet{sun2023adaptive}, the ramp metering rates are controlled at a high frequency, while the target densities of the ramp meters are controlled at a lower frequency. As another example, in the framework proposed by \citet{pasquale2017multi}, the ramp metering rates are controlled at a high frequency, whereas the vehicle splitting rates are controlled at a lower frequency. In these multi-frequency frameworks, the high-frequency control inputs must be computed sufficiently fast for real-time control, which is challenging for computationally demanding methods such as MPC, particularly in large-scale transportation networks.

The main contribution of the current paper is an innovative integrated DRL-MPC framework that divides control authority and control inputs between DRL and MPC. In the proposed framework, MPC operates at a low frequency for control measures whose slower update rate accommodates MPC's high computation demand, while DRL handles high-frequency control measures by leveraging its short deployment time. In addition, we illustrate the proposed approach on a multi-class transportation network, which, to the best of our knowledge, has not previously been considered for DRL-MPC. 

To evaluate the proposed framework, we apply it to a benchmark multi-class freeway network where the framework regulates vehicle splitting rates and ramp metering rates, and we assess its performance under noisy vehicle demands and model mismatch in the MPC prediction model. The results show that the proposed DRL-MPC framework significantly improves constraint handling compared to a fully MPC-based hierarchical controller under model mismatch, while achieving substantially faster online computation.

The rest of this paper is organized as follows: Section~\ref{sec:related_work} reviews related work on MPC and DRL for transportation network control and discusses existing hybrid DRL-MPC methods. Section~\ref{sec:DRL-MPC} presents the proposed DRL-MPC framework and its hierarchical structure, including the low-level DRL agent and the high-level MPC controller. Section~\ref{sec:Training} describes the training procedure of the DRL agent within the DRL-MPC framework, with the deterministic and stochastic actor-critic algorithms used for training detailed in \ref{app:training}. Section~\ref{sec:case_study} presents the case study on a benchmark multi-class freeway network and compares the proposed framework with benchmark controllers. Finally, Section~\ref{sec:conclusion} concludes the paper.

\section{Related work}
\label{sec:related_work}
Several control methods have been proposed to improve the efficiency and resilience of freeway and urban transportation networks in the literature \citep{majstorovic2023urban,Siri2021}. Among them, MPC and DRL have emerged as leading approaches for traffic management due to their appealing features. This section reviews relevant work on the application of MPC and DRL for transportation network control, followed by a discussion of hybrid DRL-MPC methods.

\subsection{MPC for transportation network control}

MPC has been widely adopted in transportation network control due to its capability to optimize over a prediction horizon while explicitly considering system dynamics and constraints. In freeway transportation networks, \citet{hegyi2005model} applied MPC for the optimal coordination of variable speed limits and ramp metering to reduce the TTS. \citet{deo2009model} extended this approach to multi-class freeway networks, coordinating these measures to manage different vehicle types. Similarly, \citet{liu2016model} deployed MPC in a multi-class freeway setting to simultaneously minimize the TTS, fuel consumption, and emissions, accounting for the dynamics of both cars and trucks. To address the computational burden of managing large-scale freeway networks using MPC, \citet{lin2011fast} approximated the nonlinear model of the transportation network by a mixed-integer linear model, while \citet{todorovic2020distributed} developed a distributed MPC scheme for ramp metering and variable speed limit control.

In urban transportation networks, \citet{lin2013integrated} applied MPC to reduce both the travel delays and the emissions of various exhaust gases. \citet{pham2022distributed} proposed a distributed MPC controller for urban settings to improve traffic conditions and to ensure smooth operation across roads and intersections, achieving reduced computation times relative to centralized approaches.

Despite their demonstrated performance, MPC-based approaches face notable limitations. Centralized MPC, as in \citep{deo2009model,hegyi2005model,lin2013integrated,liu2016model}, becomes computationally demanding for large-scale networks, limiting its real-time applicability. Although model simplification techniques such as approximating a nonlinear system with a mixed-integer linear formulation, as done in \citep{lin2011fast}, can reduce computation time, they introduce a trade-off between computational efficiency and control performance, since the model approximation may lead to inaccurate predictions of the transportation network dynamics. Distributed MPC strategies \citep{pham2022distributed,todorovic2020distributed} mitigate the computational burden but remain dependent on accurate prediction models, making them vulnerable to model mismatch and disturbances. Therefore, developing MPC approaches that combine computational efficiency with robustness to model errors remains a challenge in advancing transportation network control.


\subsection{DRL for transportation network control}

DRL has gained significant attention in transportation network control due to its ability to learn control policies directly from experience without requiring explicit models of the environment \citep{Haydari2022}. \citet{wang2022integrated} proposed a centralized framework that coordinates multiple ramp metering and variable speed limit controllers using an actor-critic-based DRL method to minimize the TTS, demonstrating that managing medium-scale networks with multiple on-ramps and off-ramps is feasible without complex multi-agent architectures. \citet{han2022physics} introduced a physics-informed reinforcement learning strategy that combines historical and model-generated synthetic data, enhancing learning efficiency but requiring a reliable model and large datasets for training. \citet{sun2023adaptive} developed an adaptive parameterized control scheme using deep Q-learning to tune the parameters of a state-feedback ramp metering controller online, allowing it to react to changing weather conditions. \citet{wu2020differential} proposed a DRL-based variable speed limit control approach trained with multiple reward components including the TTS, bottleneck speed, emergency braking, and emissions to improve both efficiency and safety, demonstrating generalization across different driver behaviors. \citet{jiang2025dynamic} employed a proximal policy optimization-based method for coordinated ramp metering and route guidance within a heterogeneous-agent reinforcement-learning framework, targeting lower TTS and higher mean travel speed. \citet{kolat2023multi} applied a multi-agent deep Q-learning algorithm to control traffic signals in an urban network, reducing fuel consumption and average travel time, and validated its effectiveness in a microscopic traffic simulator.

Despite their potential, DRL-based approaches face several challenges, including long training times, sample inefficiency, and the lack of mechanisms to guarantee satisfaction of state constraints, such as upper bounds on the queue lengths. Therefore, developing DRL methods that are data-efficient, constraint-aware, and capable of achieving near-optimal performance remains an active area of research for scalable and reliable transportation network control.

\subsection{Combined DRL-MPC approaches}

To exploit the complementary strengths of MPC and DRL, several hybrid frameworks have been proposed \citep{dabiri2025integrating,reiter2025synthesis}. One common approach is to train a DRL agent to adjust the MPC prediction model in order to improve model accuracy \citep{Gros2020}, as MPC relies heavily on accurate representations of system dynamics for predictions. Another widely used strategy is to parametrize the cost function and constraints of the MPC controller and to train a DRL policy to modify them, thereby enhancing MPC performance in uncertain environments \citep{hewing2020learning}. An alternative approach involves pre- or post-processing the actions generated by a DRL policy using MPC \citep{reiter2025synthesis}. In the pre-processing case, MPC serves as a reference generator for the DRL policy, whereas in post-processing, MPC acts as a safety filter, taking the DRL policy’s action as an initial guess and ensuring that the final control input satisfies the constraints. While these integration approaches have demonstrated promising performance gains over standalone DRL-based or MPC-based approaches, most existing studies do not consider hierarchical control structures in which different control inputs are updated at different frequencies, and there are fewer demonstrations on networked transportation systems, especially under multi-class dynamics. Furthermore, integrating DRL and MPC remains an emerging research area, and the approaches described above represent only a subset of the possible methods for combining these techniques.

In the context of transportation network control, \citet{sun2024novel} proposed a novel hierarchical DRL-MPC framework. In this framework, a high-level MPC component operates at a low frequency to provide a baseline control input, while a DRL component operates at a high frequency to modify the output of MPC in real time. Simulation results demonstrated that this framework outperformed standalone MPC and DRL methods in terms of TTS and constraint satisfaction in the presence of model uncertainties and external disturbances. 

\Citet{sun2024adaptive} proposed another framework that integrates DRL and MPC to address different sources of model mismatches. This framework uses DRL to adapt all components of a parametrized MPC scheme (e.g., the objective function, state-feedback control law, optimization settings, and system model) to manage model mismatches caused by changing environments and disturbances. In a freeway network control case study, the framework’s ability to adjust state-feedback control laws under model mismatches and varying weather conditions was tested. Simulation results showed that the proposed framework reduced computational complexity relative to conventional MPC while achieving better control performance than DRL-based controllers. 

Subsequently, \citet{airaldi2025reinforcement} used a DRL policy to adjust selected parameters of the prediction model, objective function, and cost function of the MPC controller to control ramp metering in freeway networks. Simulation results demonstrated that their approach improved traffic-control performance and satisfied constraints even when the MPC controller used a mismatched prediction model and initially poorly tuned parameters. 

Despite the promising results of these approaches \citep{airaldi2025reinforcement,sun2024adaptive,sun2024novel}, their focus has largely been on adapting MPC outputs and parameters via DRL to address model uncertainties in single-class transportation networks. As such, the potential of DRL-MPC integration for multi-class transportation networks in settings where different control measures are updated at different frequencies remains underexplored, leaving opportunities for further research.

\section{The DRL-MPC framework} \label{sec:DRL-MPC}

\begin{figure}[tbp]
\centering
\includegraphics[width=0.95\textwidth]{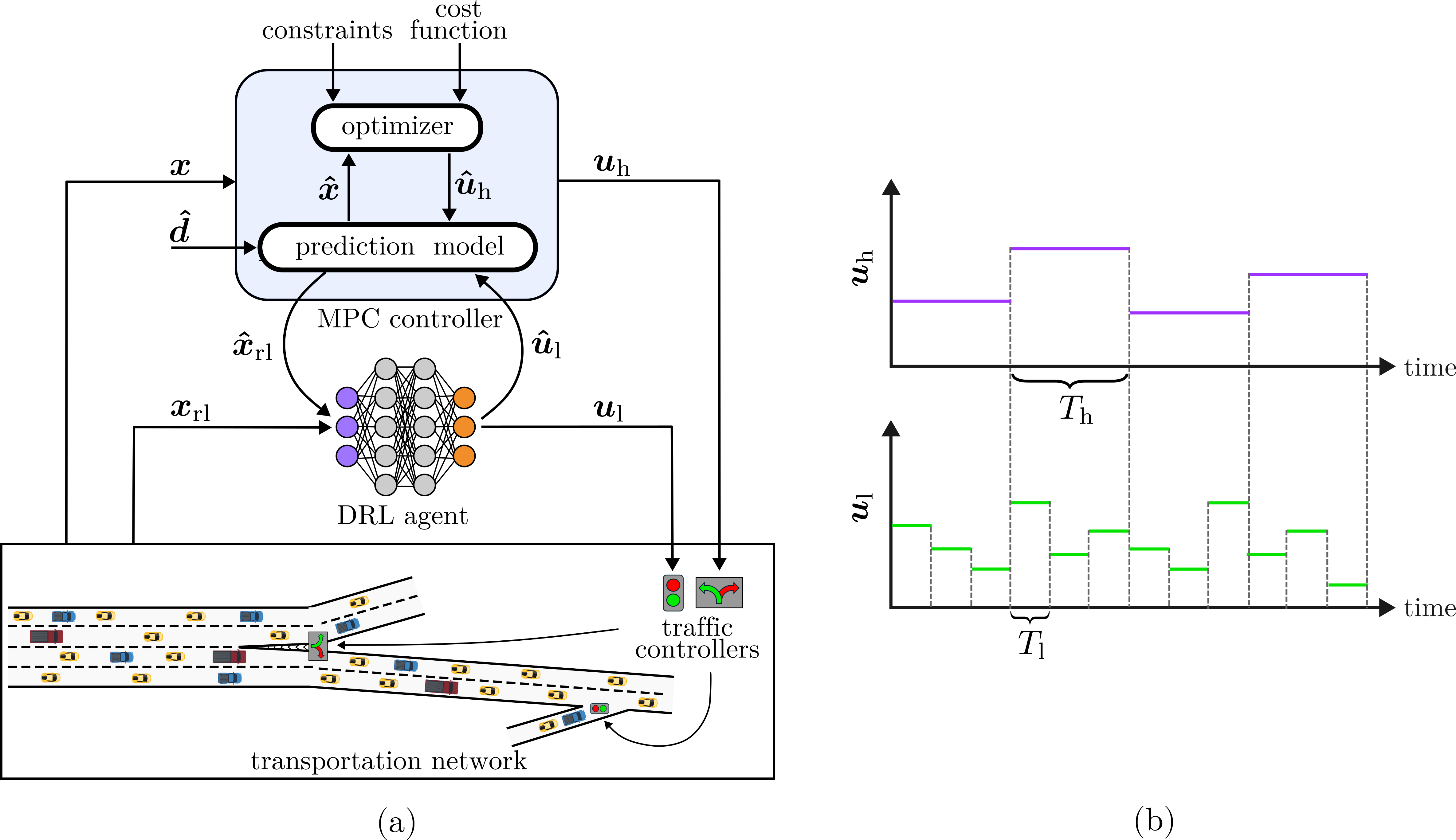}
\caption{(a) Block diagram illustrating the proposed DRL-MPC framework that divides control inputs between the DRL agent and the MPC controller. (b) Time scales for the low-frequency control inputs of the high-level MPC controller and the high-frequency control inputs of the low-level DRL agent. 
} \label{RL_MPC_simple}
\end{figure}

The proposed DRL-MPC framework divides control inputs between DRL and MPC. The framework has a hierarchical structure where the MPC controller operates at the high level, running at a low frequency, and the DRL agent operates at the low level, running at a high frequency (see Figure~\ref{RL_MPC_simple}). Consequently, the DRL agent can compute high-frequency control inputs quickly due to its fast online deployment capabilities, whereas the MPC controller has sufficient time to calculate low-frequency control inputs.

The framework aims to minimize a given objective function, such as the TTS in a transportation network. Accordingly, as explained in the following section, the MPC cost function and the DRL reward function are designed to complement each other in achieving a common objective. Moreover, constraint violations, such as exceeding the vehicle queue length limits, are incorporated into the DRL reward function as penalty terms, allowing the DRL agent to learn to take actions that aim to respect the constraints imposed by the MPC controller.

Note that in the framework, the MPC controller uses the trained DRL agent, whose policy is held fixed during deployment, to obtain high-frequency control inputs required by MPC for predicting the future states of the transportation network (see Figure~\ref{RL_MPC_simple}(a)). This allows the MPC controller to predict the future states of the transportation network over its prediction window, even though the control input of the lower level is not determined by the MPC controller but determined by the DRL agent.

In the following sections, the DRL agent, the MPC controller, and the training methods for the DRL-MPC framework are explained in detail. The key notations used for the DRL-MPC framework are summarized in Table~\ref{tab:nomenclature}. 

\begin{table}[tbp]
\centering
\caption{Key notations used for the DRL-MPC framework}
\label{tab:nomenclature}
\small
\begin{tabular}{ll}
\toprule
\textbf{Notation} & \textbf{Definition} \\
\midrule
$k$ & Network sampling step counter \\
$k_\mathrm{l}$ & Low-level (DRL) control step counter \\
$k_\mathrm{h}$ & High-level (MPC) control step counter \\
$T$ & Network sampling time interval \\
$T_\mathrm{l}$ & Low-level control sampling time interval\\
$T_\mathrm{h}$ & High-level control sampling time interval\\
$m_\mathrm{l}$ & Ratio of low-level control and network sampling time intervals \\
$m_\mathrm{h}$ & Ratio of high-level control and network sampling time intervals \\
$N_\mathrm{p}$ & Prediction horizon (in high-level steps) \\
$\boldsymbol{x}(k)$ & State vector of the transportation network at sampling step $k$ \\
$\boldsymbol{x}_\mathrm{rl}(k_\mathrm{l})$ & State vector of the DRL agent at control step $k_\mathrm{l}$ \\
$\boldsymbol{u}_\mathrm{l}(k_\mathrm{l})$ & Low-level control input computed at control step $k_\mathrm{l}$ \\
$\boldsymbol{\bar{u}}_\mathrm{l}(k)$ & Low-level control input applied at network sampling step $k$ \\
$\boldsymbol{u}_\mathrm{h}(k_\mathrm{h})$ & High-level control input computed at control step $k_\mathrm{h}$ \\
$\boldsymbol{\bar{u}}_\mathrm{h}(k)$ & High-level control input applied at network sampling step $k$ \\
$\boldsymbol{\hat{x}}(k)$ & Predicted future network state at sampling step $k$ \\
$M$ & Total number of training episodes \\
$k^{\text{max}}_{\text{eps}}$ & Maximum simulation steps per episode \\
$\pi_\theta$ & DRL actor policy network with parameters $\theta$ \\
$Q_\phi$ & DRL critic value network with parameters $\phi$ \\
$\mathcal{D}$ & Experience replay buffer \\
$n$ & Number of steps to look ahead for the reward \\
$r(k_\mathrm{l})$ & Reward received at control step $k_\mathrm{l}$ \\
$\gamma$ & Discount factor \\
$\tau$ & Smoothing factor for Polyak averaging \\
\bottomrule
\end{tabular}
\end{table}

\subsection{Low-level DRL agent}
\label{subsec:drl_agent}

In the proposed framework, the DRL agent determines the high-frequency control inputs, such as ramp metering rates in a freeway network \citep{pasquale2017multi}. Let $k_\mathrm{l}$ represent the control step counter for the low-level DRL agent, with a control time interval of length $T_\mathrm{l}$. The DRL agent receives the network states, $\boldsymbol{x}_\mathrm{rl}(k_\mathrm{l})$, at every step $k_\mathrm{l}$. The states of the DRL agent are defined as
\begin{equation}\label{eq:DRL_states}
\boldsymbol{x}_\mathrm{rl}(k_\mathrm{l}) = [\boldsymbol{\bar{x}}^\top (k_\mathrm{l}),\boldsymbol{\bar{d}}^\top (k_\mathrm{l})]^\top,
\end{equation}
where $\boldsymbol{\bar{x}}(k_\mathrm{l})$ and $\boldsymbol{\bar{d}}(k_\mathrm{l})$ are the normalized states of the transportation network (e.g., densities, outflows, and velocities of vehicles on road segments) and the normalized current traffic demands, respectively. Normalization is applied by scaling the network states and traffic demands by constant gains to bring these variables to similar orders of magnitude, thereby enhancing the efficiency of the learning process.

The relationship between the DRL control sampling time interval and the network sampling time interval is described as
\begin{equation*}
T_\mathrm{l} = m_\mathrm{l} T, \quad k = m_\mathrm{l} k_\mathrm{l}, \quad m_\mathrm{l} \in \mathbb{N}^{+}, \quad m_\mathrm{l} \geq 1,
\end{equation*}
where $k$ is the network sampling step counter, $T$ is the network sampling time interval length, and $\mathbb{N}^{+}$ denotes the set of positive integers.

The DRL agent computes the high-frequency control inputs every $T_\mathrm{l}$ time units as

\begin{equation*}
\boldsymbol{u}_\mathrm{l}(k_\mathrm{l}) \sim \pi_\theta\big(\cdot \mid \boldsymbol{x}_\mathrm{rl}(k_\mathrm{l})\big), \quad \boldsymbol{u}_\mathrm{l}(k_\mathrm{l}) \in \mathcal{U}_\mathrm{l},
\end{equation*}
where $\pi_\theta$ is the DRL policy network with parameters $\theta$, $\mathcal{U}_\mathrm{l}$ is the constraint set for the high-frequency control inputs, and $\boldsymbol{u}_\mathrm{l} \sim \pi_\theta(\cdot \mid \cdot)$ implies sampling from a policy distribution\footnote{In the case of a deterministic policy, this corresponds to sampling from a Dirac delta distribution where the entire probability mass is concentrated at the single output value of the policy network.}. Then, the high-frequency control inputs that are applied to the transportation network during the control sampling time interval $T_\mathrm{l}$ are calculated as

\begin{equation*}
\boldsymbol{\bar{u}}_\mathrm{l}(k) = \boldsymbol{u}_\mathrm{l}(k_\mathrm{l}), \quad k \in \{ m_\mathrm{l} k_\mathrm{l}, m_\mathrm{l} k_\mathrm{l}+1, ..., m_\mathrm{l} k_\mathrm{l} + m_\mathrm{l} - 1 \},
\end{equation*}
meaning that the control input $\boldsymbol{u}_\mathrm{l}(k_\mathrm{l})$ is applied over a time interval of duration $T_\mathrm{l}$.


\subsection{High-level MPC controller}
\label{subsec:mpc_controller}

\begin{figure}[tbp]
\centering
\includegraphics[width=0.7\textwidth]{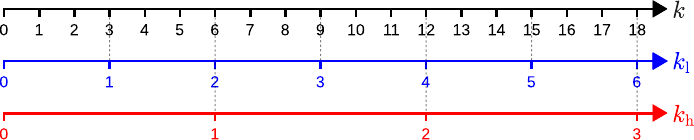}
\caption{Relationship between the network sampling step counter $k$, low-level control step counter $k_\mathrm{l}$, and high-level control step counter $k_\mathrm{h}$ for $m_\mathrm{l} = 3$ and $m_\mathrm{h} = 6$.} \label{fig:step_counters}
\end{figure}

The MPC controller computes the low-frequency control inputs, such as vehicle splitting rates in a freeway network \citep{pasquale2017multi}. The MPC controller employs a receding-horizon strategy by computing the control sequence
\begin{equation*}
\boldsymbol{\tilde{u}}_\mathrm{h}(k_\mathrm{h}) = \left[ \boldsymbol{u}_\mathrm{h}^\top(k_\mathrm{h}), \boldsymbol{u}^\top_\mathrm{h}(k_\mathrm{h} + 1), \dots, \boldsymbol{u}^\top_\mathrm{h}(k_\mathrm{h} + N_\mathrm{p} - 1) \right]^\top
\end{equation*}
over a prediction horizon of \( N_\mathrm{p} \) high-level time steps, each of duration $T_\mathrm{h}$, with a step counter $k_\mathrm{h}$, where
\begin{equation*}
T_\mathrm{h} = m_\mathrm{h} T, \quad 
k = m_\mathrm{h} k_\mathrm{h}, \quad 
m_\mathrm{h} \in \mathbb{N}^{+}, \; m_\mathrm{h} \geq m_\mathrm{l}.
\end{equation*}
To ensure temporal synchronization, $m_\mathrm{h}$ is selected as an integer multiple\footnote{Setting $m_\mathrm{h}=m_\mathrm{l}$ is also possible and the framework remains applicable; in this case, however, the hierarchical structure (in which the MPC controller runs at a low frequency and the DRL agent runs at a high frequency) is no longer present.} of $m_\mathrm{l}$, guaranteeing that MPC and DRL control input updates align (see Figure~\ref{fig:step_counters}).

The predicted future states of the freeway network are denoted as
\begin{equation*}
\boldsymbol{\tilde{x}}(k_\mathrm{h}) = \left[ \boldsymbol{\hat{x}}^\top(m_\mathrm{h}k_\mathrm{h} + 1), \boldsymbol{\hat{x}}^\top(m_\mathrm{h}k_\mathrm{h} + 2), \dots, \boldsymbol{\hat{x}}^\top(m_\mathrm{h} k_\mathrm{h}+ m_\mathrm{h} N_\mathrm{p}) \right]^\top,
\end{equation*}
where $\boldsymbol{\hat{x}}(k)$ is the predicted future state of the freeway network at sampling step $k$. The states are predicted at the network sampling steps to evaluate the objective function accurately, even though the control inputs are updated only at the high-level control steps.

The MPC controller computes the control inputs over the prediction horizon by solving the following optimization problem:
\begin{subequations}\label{eq:MPC}
\begin{align}
\min_{\boldsymbol{\tilde{u}}_\mathrm{h}(k_\mathrm{h}),\boldsymbol{\tilde{x}}(k_\mathrm{h})} \;\; 
& \sum^{m_\mathrm{h} N_\mathrm{p}}_{l=1} J_\mathrm{TTS} \big( \boldsymbol{\hat{x}}(m_\mathrm{h} k_\mathrm{h} + l)\big) \nonumber\\
&+ \omega_\mathrm{u_h}
\sum_{j=0}^{N_\mathrm{p}-1}
\lVert \boldsymbol{u}_\mathrm{h}(k_\mathrm{h}+j) - \boldsymbol{u}_\mathrm{h}(k_\mathrm{h}+j-1) \rVert_2^2  \nonumber \\
\text{s.t.}\;\;
& \boldsymbol{\hat{x}}(m_\mathrm{h} k_\mathrm{h} + l + 1) = \nonumber \\
&\indent F\big(\boldsymbol{\hat{x}}(m_\mathrm{h} k_\mathrm{h} + l ),\boldsymbol{\hat{u}}_\mathrm{h}(m_\mathrm{h} k_\mathrm{h} + l), \boldsymbol{\hat{u}}_\mathrm{l}(m_\mathrm{h} k_\mathrm{h} + l), \boldsymbol{\hat{d}}(m_\mathrm{h} k_\mathrm{h} + l)\big), \nonumber\\
&\indent \indent \text{for } l \in \{0,1,\dots,m_\mathrm{h}N_\mathrm{p}-1 \}, \label{eq:MPC_dyn} \\
& \boldsymbol{\hat{x}}(m_\mathrm{h} k_\mathrm{h}) = \boldsymbol{x}(m_\mathrm{h} k_\mathrm{h}), \label{eq:MPC_init} \\
& \boldsymbol{\hat{u}}_\mathrm{h}(m_\mathrm{h} (k_\mathrm{h} +o) + l) = \boldsymbol{u}_\mathrm{h}(k_\mathrm{h} + o), \nonumber\\
&\indent \text{for } l \in \{0,1,\dots,m_\mathrm{h}-1 \}, \; o \in \{0,1,\dots,N_\mathrm{p}-1\}, \label{eq:MPC_hold_h} \\
& \boldsymbol{\hat{u}}_\mathrm{l}(m_\mathrm{l} (k_\mathrm{l} +o) + l) \sim  \pi_\theta \big(\cdot \mid \boldsymbol{\hat{x}}_\mathrm{rl}(k_\mathrm{l} + o)\big), \nonumber\\
&\indent \text{for } l \in \{0,1,\dots,m_\mathrm{l}-1 \}, \; o \in \{0,1,\dots,\frac{m_\mathrm{h}}{m_\mathrm{l}}N_\mathrm{p}-1\}, \label{eq:MPC_policy_l} \\
& \boldsymbol{\hat{x}}(m_\mathrm{h}k_\mathrm{h}+l) \in \mathcal{X}, \quad l \in \{1,\dots,m_\mathrm{h}N_\mathrm{p}\}, \nonumber\\
& \boldsymbol{u}_\mathrm{h}(k_\mathrm{h}+j) \in \mathcal{U}_\mathrm{h}, \quad j \in \{0,\dots,N_\mathrm{p}-1\}. \label{eq:MPC_constraints}
\end{align}
\end{subequations}
In \eqref{eq:MPC}, the first term in the objective function, $J_\mathrm{TTS}$, represents the TTS, and the second term represents the quadratic penalty on fluctuations between consecutive high-level control inputs, with penalty weight $\omega_\mathrm{u_h}$, where $\lVert \cdot \rVert_2$ denotes the Euclidean norm. Constraint \eqref{eq:MPC_dyn} represents the prediction model dynamics, where $F$ represents the multi-class METANET model (see \ref{A1:Metanet}) and $\boldsymbol{\hat{d}}(k)$ denotes the estimated traffic demands. Constraint \eqref{eq:MPC_init} sets the initial condition of the predicted state trajectory to the measured state at the current high-level step. Constraint \eqref{eq:MPC_hold_h} maps the MPC control inputs from high-level to low-level control steps using a zero-order hold strategy. Constraint \eqref{eq:MPC_policy_l} specifies that the low-level control inputs used in the prediction are generated by the DRL policy $\pi_\theta$ based on the predicted DRL agent state $\boldsymbol{\hat{x}}_\mathrm{rl}(k)$ (see Figure~\ref{RL_MPC_simple}(a)). Finally, \eqref{eq:MPC_constraints} collects the state and high-level input constraints, where $\mathcal{X}$ and $\mathcal{U}_\mathrm{h}$ denote the corresponding constraint sets, respectively.

After solving \eqref{eq:MPC}, the MPC controller applies only the first control input \( \boldsymbol{u}_\mathrm{h}(k_\mathrm{h}) \) to the system during the high-level control step, following a receding-horizon strategy:
\begin{equation*}
    \boldsymbol{\bar{u}}_\mathrm{h}(k) = \boldsymbol{u}_\mathrm{h}(k_\mathrm{h}), \quad k \in \{ m_\mathrm{h}k_\mathrm{h},  m_\mathrm{h}k_\mathrm{h}+1, ...,  m_\mathrm{h}k_\mathrm{h} + m_\mathrm{h} - 1 \},
\end{equation*}
and repeats this procedure at every $T_\mathrm{h}$ time units.

Note that \eqref{eq:MPC} leads to a nonlinear and nonconvex optimization problem for the multi-class METANET model \citep{liu2016model}, which necessitates the use of a nonlinear solver such as multi-start sequential quadratic programming, simulated annealing, or genetic algorithms \citep{floudas2013state}.

\section{Training the DRL-MPC framework}\label{sec:Training}

\begin{figure}[tbp]
\centering
\includegraphics[width=0.65\textwidth]{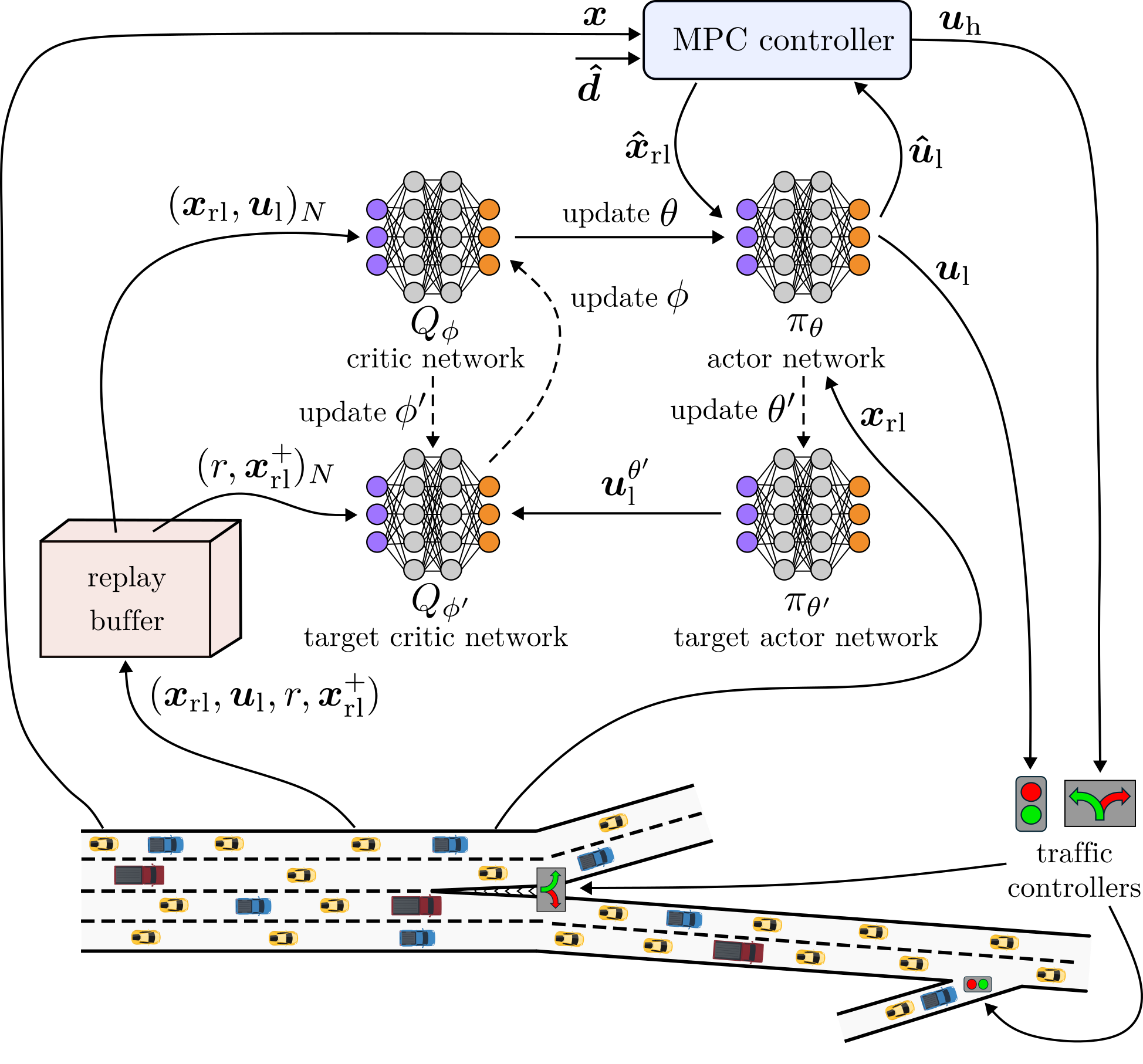}
\caption{Block diagram illustrating the training process of the DRL-MPC framework using the DDPG algorithm.
} \label{RL_MPC_fig}
\end{figure}

We train the DRL component of the DRL-MPC framework using data samples collected while the DRL agent operates at the low level together with the MPC controller operating at the high level in the DRL-MPC architecture (see Figure~\ref{RL_MPC_fig}). The collected data samples are then used to update the DRL policy, so that the learned policy accounts for the effect of the MPC controller within the DRL-MPC framework.

We consider two actor-critic algorithms for training the DRL agent within the DRL-MPC framework: Deep Deterministic Policy Gradient (DDPG) \citep{lillicrap2015continuous} and Soft Actor-Critic (SAC) \citep{haarnoja2018soft}, given their success in many control benchmarks \citep{wang2022deep}, including freeway traffic control \citep{airaldi2025reinforcement,lu2025soft}. DDPG learns a deterministic policy, whereas SAC learns a stochastic policy. This allows us to compare the performance of the proposed DRL-MPC framework when the DRL agent is trained using either a deterministic or a stochastic actor-critic algorithm.

Both DDPG and SAC utilize off-policy learning, which enhances sample efficiency by allowing the reuse of past experience tuples $(\boldsymbol{x}_\mathrm{rl},\boldsymbol{u}_\mathrm{l},r,\boldsymbol{x}^{+}_\mathrm{rl})$ stored in a replay buffer $\mathcal{D}$, where $r$ is the collected reward and $\boldsymbol{x}^{+}_\mathrm{rl}$ is the next observed state after taking the action $\boldsymbol{u}_\mathrm{l}$. This sample efficiency is particularly beneficial for the proposed DRL-MPC framework, as training involves computing high-level control inputs through MPC, which increases simulation time and, consequently, the time required for each sample. In both algorithms, the following reward function is used:
\begin{equation} \label{eq:RLreward}
\begin{aligned}
        r(k_\mathrm{l}) = -\Bigg(
        \sum^{m_\mathrm{l}}_{l=1} J_\mathrm{TTS}\big( \boldsymbol{x}(m_\mathrm{l} k_\mathrm{l} + l)\big)
        + \omega_\mathrm{u_l} \lVert \boldsymbol{u}_\mathrm{l}(k_\mathrm{l}) - \boldsymbol{u}_\mathrm{l}(k_\mathrm{l}-1) \rVert_2^2
        + P^\mathrm{rl}( k_\mathrm{l})
        \Bigg)
\end{aligned}
\end{equation}
where the first term coincides with the TTS term used by the MPC controller in \eqref{eq:MPC}, and the second term represents the quadratic penalty on fluctuations between consecutive low-level control inputs, with penalty weight $\omega_\mathrm{u_l}$. The term $P^\mathrm{rl}$ is a positive state constraint penalty term used to penalize violations of the state constraints in \eqref{eq:MPC_constraints}, such as queue length constraints at the mainstream and on-ramp origins, since the DDPG and SAC implementations used in this study do not directly enforce hard state constraints. Note that the reward is defined as the negative of the TTS and penalty terms, so that maximizing the cumulative reward in DRL corresponds to minimizing these terms, aligning the objective of the DRL policy with that of the MPC controller.

The overall training procedure is summarized in Algorithm~\ref{alg:training}, and the detailed formulations of how DDPG and SAC are utilized for training the DRL-MPC framework are provided in \ref{app:training}.

\section{Case study}\label{sec:case_study}

The proposed DRL-MPC framework is evaluated in a case study for freeway traffic control on a multi-class network with two vehicle classes and two control measures, where a vehicle splitting rate (via a variable message sign for route guidance) is updated at a low frequency and two ramp metering rates are updated at a higher frequency. The performance of the DRL-MPC framework is compared with that of a hierarchical MPC controller, a state-feedback-MPC controller, and the no-control case. Robustness of the controllers is assessed under prediction model mismatch in the MPC prediction models and under noisy demand profiles.

The remainder of this section is organized as follows: Section~\ref{subsec:case_study_setup} describes the experimental setup and simulation scenarios, Section~\ref{subsec:case_study_controllers} summarizes the controllers and evaluation metrics, Section~\ref{subsec:training_results} reports training results, and Section~\ref{subsec:deployment_results} presents the case study results and compares the controllers based on the evaluation metrics. The source code for this case study is available at \url{https://github.com/GirayOnur/Sharing-Control-Authority-Between-DRL-and-MPC}.

\subsection{Experimental setup}\label{subsec:case_study_setup}

\begin{figure}[tbp]
\centering
\includegraphics[width=0.65\textwidth]{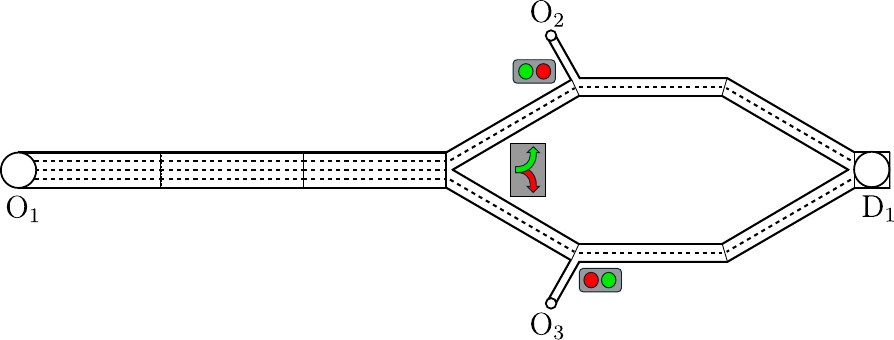}
\caption{The benchmark freeway network with one variable message sign for route guidance and two metered on-ramps.}
\label{fig:network_layout}
\end{figure}

We consider a benchmark freeway network (see Figure~\ref{fig:network_layout}) with two vehicle classes. The network consists of a mainstream stretch of three segments with four lanes and a mainstream origin $O_1$, and the mainstream stretch splits into two routes: a primary route with on-ramp origin $O_2$ and a secondary route with on-ramp origin $O_3$. Each route consists of three segments with two lanes, and both routes merge at the single destination $D_1$. Each on-ramp has one lane. The free-flow capacity of all origins is $2000$~veh/h/lane, and the queue length limits are $200$ vehicles at $O_1$ and $100$ vehicles at each on-ramp origin. Each segment is $1$~km long, the network sampling time interval is $10$~s, and the case study simulation duration is $2.5$~h.

\begin{figure}[tbp]
\centering
\includegraphics[width=0.85\textwidth]{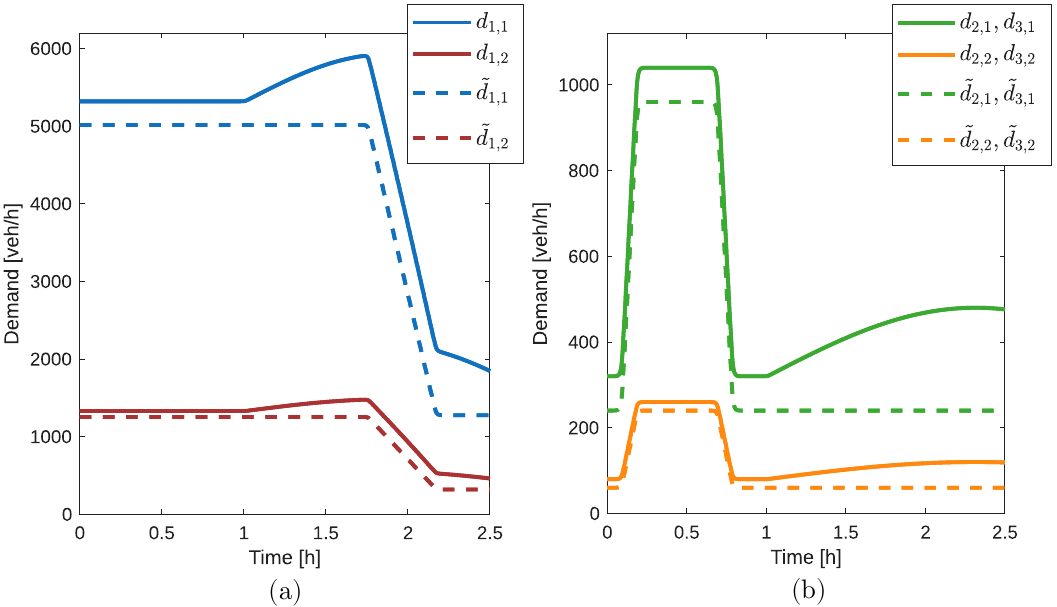}
\caption{The vehicle demand profiles used in the case study for the (a) mainstream origin and (b) on-ramp origins. The solid curves show the nominal demands, and the dashed curves show the estimated demands used for MPC predictions in the mismatch cases.} \label{fig:demands}
\end{figure}

The multi-class transportation network shown as the controlled system in Figure~\ref{RL_MPC_simple}(a) is simulated using the multi-class METANET model (see \ref{A1:Metanet}). The parameter values used for simulations and the perturbed parameter values used by MPC under model mismatch are reported in Table~\ref{table:metanet_params}. The nominal demand profiles and the estimated demands used for MPC predictions in the mismatched cases are shown in Figure~\ref{fig:demands}. These demand profiles are selected to induce severe congestion in the absence of effective ramp metering and vehicle splitting rate control, providing a challenging operating condition for evaluating traffic controller performance. To evaluate robustness to demand variability, we generate noisy demand profiles by adding zero-mean Gaussian noise to the nominal demand profiles, with origin- and class-specific standard deviations:
\begin{equation*}
(\sigma_{1,1},\sigma_{1,2},\sigma_{2,1},\sigma_{2,2},\sigma_{3,1},\sigma_{3,2})
=
(200,50,40,10,40,10),
\end{equation*}
and then smoothing the resulting demands using a third-order low-pass Butterworth filter with a normalized cutoff frequency of $0.1$.

\begin{table}[tbp]
\centering
\begin{threeparttable}
\caption{Original and perturbed values of multi-class freeway network parameters in the case study.}
\label{table:metanet_params}
\small
\begin{tabular}{l c c c}
\toprule
\textbf{Parameter} & \textbf{Unit} & \textbf{Original Value} & \textbf{Perturbed Value} \\
\midrule
$T$ & [s]\tnote{a} & 10 & 10 \\
$\tau_c$ & [s]\tnote{a} & 18 & 18 \\
$\kappa_c$ & [veh/km/lane] & 40 & 40 \\
$\eta_c$ & [km$^2$/h] & 60 & 60 \\
$L_{m}$ & [km] & 1 & 1 \\
$L^\mathrm{veh}_{c}$ & [m] & 5 & 5 \\
$\sigma_c$ & - & 0.0122 & 0.0122 \\
$a_{m,1}$ & - & 1.800 & 1.950 \\
$a_{m,2}$ & - & 2.023 & 2.317 \\
$v^{\mathrm{free}}_{m,1}$ & [km/h] & 110 & 102 \\
$v^{\mathrm{free}}_{m,2}$ & [km/h] & 83.33 & 75.33 \\
$\rho^{\mathrm{max}}_{m}$ & [veh/km/lane] & 180 & 165 \\
$\rho^{\mathrm{crit}}_{m}$ & [veh/km/lane] & 33.5 & 38.5 \\
\bottomrule
\end{tabular}
\begin{tablenotes}[flushleft]
\footnotesize
\item[a] The tabulated values of $T$ and $\tau_c$ are expressed in seconds and converted to hours when used in the METANET equations.
\end{tablenotes}
\end{threeparttable}
\end{table}

We evaluate all controllers under four scenarios that combine demand noise and prediction model mismatch. The scenarios are defined as follows:
\begin{itemize}
    \item \textit{Scenario 1:} The simulator of the multi-class network uses the original values of the parameters and the nominal demand profiles. The MPC-based controllers use the original values of the parameters in their prediction model and assume the nominal demands as demand predictions.
    \item \textit{Scenario 2:} The simulator of the multi-class network uses the original values of the parameters and noisy demand profiles. The MPC-based controllers use the original values of the parameters in their prediction model and assume the noisy demands as demand predictions.
    \item \textit{Scenario 3:} The simulator of the multi-class network uses the original values of the parameters and the nominal demand profiles as in Scenario~1. The MPC-based controllers use the perturbed values of the parameters in their prediction model and assume the estimated demands as demand predictions.
    \item \textit{Scenario 4:} The simulator of the multi-class network uses the original values of the parameters and noisy demand profiles as in Scenario~2, while the MPC-based controllers use the perturbed values of the parameters in their prediction model and assume the estimated demands as demand predictions as in Scenario~3.
\end{itemize}

To ensure that the scenario starts from a congested operating condition, we first simulate the network for $10$~min under the no-control setting, with a fixed vehicle splitting rate of $0.5$ and the ramp metering rates set to $1$. All controllers are activated after this initialization period, and performance metrics are evaluated over the remaining controlled interval of $2.5$~h. The vehicle splitting rate and the two ramp metering rates are bounded in the interval $[0,1]$. The ramp metering rates are updated every $T_\mathrm{l}=60$~s, while the vehicle splitting rate is updated every $T_\mathrm{h}=300$~s. The same vehicle splitting rate and ramp metering rates are applied to both vehicle classes.

In each scenario, the hierarchical MPC framework and the no-control case are each evaluated $10$ times. SF-MPC is evaluated $10$ times for each of five independently tuned PI-ALINEA parameter sets, while DDPG-MPC and SAC-MPC are each evaluated $10$ times for each of five independently trained agents. This yields $50$ evaluations each for SF-MPC, DDPG-MPC, and SAC-MPC per scenario. For each of the $10$ evaluation runs, all control frameworks use the same random seed. This seed is used to generate the multi-start initializations for the MPC-based frameworks. In Scenarios~2 and~4, the same seed is also used to generate the noisy demand profiles.

All case study evaluations are performed in Matlab R2025a on a PC with a 12th Gen Intel Core i3-12100F CPU and 16~GB RAM.

\subsection{Compared control frameworks}\label{subsec:case_study_controllers}

The proposed DRL-MPC framework is evaluated against a hierarchical MPC controller, a state-feedback-MPC controller, and a no-control case, with the details of each approach described below. All controllers are evaluated on the same freeway network, have access to the same network state measurements, and are subject to identical control input bounds.

\subsubsection{DRL-MPC framework}\label{subsubsec:drl_mpc_case}

The DRL-MPC framework assigns vehicle splitting rate control to the MPC controller and ramp metering rate control to the DRL agent.

MPC computes the control inputs by solving the optimization problem \eqref{eq:MPC} using the Matlab \texttt{fmincon} function with the sequential quadratic programming (SQP) algorithm. A multi-start strategy with $5$ initializations is used to balance the trade-off between control performance and computation time. To achieve sufficient numerical accuracy while maintaining computational efficiency, the optimality tolerance, step tolerance, and constraint tolerance are set to $10^{-2}$.

While training the DRL agent, the MPC controller uses a soft penalty for queue length limit violations instead of hard constraints. This choice avoids infeasibility at early training stages, when the DRL policy may not yet regulate the ramp metering rates effectively and the MPC controller may not be able to compute a split rate that satisfies the queue length limits. Specifically, the penalty
\begin{equation}\label{eq:soft_pen_MPC}
P^\mathrm{mpc} (k) = \omega_\mathrm{w} \sum_{o\in\{O_1,O_2,O_3\}}
\Bigl(\max\{0,\,w_o(k)-w_o^{\max}\}\Bigr)^2
\end{equation}
is included as an additional term in the objective function of \eqref{eq:MPC}, where $\omega_\mathrm{w}$ in \eqref{eq:soft_pen_MPC} is the queue length penalty weight, $w_o(k)=\sum_{c=1}^{C}w_{o,c}(k)$ is the total queue length at network sampling step $k$ at origin $o$, and $w_o^{\max}$ is the queue length limit at origin $o$. Although the soft penalty term in \eqref{eq:soft_pen_MPC} is used during training of the DRL agent, during deployment in the case study evaluation the MPC controller enforces queue length limits as hard constraints in \eqref{eq:MPC}. 
In \eqref{eq:MPC}, the queue length penalty weight is set to $\omega_\mathrm{w}=1$ and the vehicle splitting rate smoothing weight to $\omega_\mathrm{u_h}=2.0$, with $m_\mathrm{h}=30$ and $m_\mathrm{l}=6$ corresponding to the update intervals $T_\mathrm{h}=300$ s and $T_\mathrm{l}=60$ s relative to the network sampling time interval of 10 s. The MPC controller uses a prediction horizon\footnote{Although small in the number of high-level steps, this horizon spans $600$~s due to the long high-level control sampling time interval, $T_\mathrm{h}$, corresponding to $10$ low-level control steps or $60$ network sampling steps.} $N_{\mathrm{p}}=2$.

During training, the DRL reward is calculated using \eqref{eq:RLreward} with $\omega_{\mathrm{u_l}}=0.4$, obtained by scaling $\omega_{\mathrm{u_h}}$ according to the ratio between $m_\mathrm{l}$ and $m_\mathrm{h}$. The reward in \eqref{eq:RLreward} is multiplied by a constant scaling factor of $1/30$ before being stored in the replay buffer, as scaling the reward signal has been shown to improve learning performance and stability of off-policy actor-critic algorithms \citep{haarnoja2018soft}. The queue violation term $P^\mathrm{rl}$ in \eqref{eq:RLreward} is implemented using the maximum total queue length within each low-level control step. Let $\bar{w}_o(k_\mathrm{l})=\max_{l\in\{1,\dots,m_\mathrm{l}\}} w_o(m_\mathrm{l}k_\mathrm{l}+l)$ denote the maximum total queue length at origin $o\in\{O_1,O_2,O_3\}$ over $T_\mathrm{l}$. Then,
\begin{equation}
P^\mathrm{rl}(k_\mathrm{l})=\sum_{o\in\{O_1,O_2,O_3\}}
\left(p_\mathrm{s}+\frac{\bar{w}_o(k_\mathrm{l})}{w_\mathrm{s}}\right)\,
\mathbb{I}\{\bar{w}_o(k_\mathrm{l})>w_o^{\max}\},
\end{equation}
where $\mathbb{I}\{\cdot\}$ is the indicator function (i.e., it equals $1$ when its condition holds and $0$ otherwise), $p_\mathrm{s}=10$ is a constant violation penalty, and $w_\mathrm{s}=100$ is a queue scaling factor.

The DRL agent is trained using the DDPG and SAC algorithms described in \ref{subsec:train_ddpg} and~\ref{subsec:train_sac}, respectively. Both DRL variants use the same observation and action definitions. The observation is an $81$-dimensional vector comprising the following values: for each of the nine road segments, two class-specific mean speeds, two class-specific densities, the total density, and two class-specific outflows; and for each of the three origins, two class-specific queue lengths, two class-specific outflows, and two current class-specific demands. The DRL action is a $2$-dimensional vector corresponding to the two ramp metering rates. The observation is normalized by a constant scaling vector selected according to the magnitude of each component to improve training stability: speed, density, and queue-related states are normalized by dividing by $100$, and flow-related states and demand values are normalized by dividing by $1000$.

The network architectures and the training hyperparameters used for the DDPG and SAC agents in the case study are reported in \ref{subsec:hyperparams}.

We run the training in Matlab R2025a on the DelftBlue supercomputer \citep{DHPC2024}, using 32 CPU cores with $2100$~MB memory per CPU core, and asynchronous parallel training with the Matlab Reinforcement Learning Toolbox. For each scenario defined in Section~\ref{subsec:case_study_setup}, five independent agents are trained for DDPG and five for SAC.

\subsubsection{Hierarchical MPC framework}\label{subsubsec:hm_mpc_case}
The hierarchical MPC controller consists of two coupled MPC controllers operating at different time scales: a high-level MPC controller that computes the vehicle splitting rates and a low-level MPC controller that computes the ramp metering rates. The high-level MPC controller uses the same formulation and parameter settings as in Section~\ref{subsubsec:drl_mpc_case}, but differs in how low-level inputs are treated within the prediction process. In the proposed DRL-MPC framework, predicted ramp metering inputs are generated by evaluating the DRL policy during the high-level MPC prediction. In the hierarchical MPC controller, the high-level MPC prediction uses the most recent ramp metering trajectory computed by the low-level MPC controller, shifted by one step with the terminal value duplicated, to predict the effect of ramp metering, thereby avoiding the prohibitive computational cost of solving additional low-level MPC problems inside the high-level MPC problem.

The low-level MPC controller computes the ramp metering rates $\boldsymbol{u}_\mathrm{l}$. The controller uses $m_\mathrm{l}=6$ simulation steps per low-level control step, matching the low-level control step of the DRL-MPC framework in Section~\ref{subsubsec:drl_mpc_case}, and a low-level prediction horizon $N_{\mathrm{p,l}}=10$, defined in low-level control steps and chosen so that the low-level prediction spans the same time interval as the high-level MPC prediction. Its objective is similar to \eqref{eq:MPC}, but the smoothing term penalizes squared changes in $\boldsymbol{u}_\mathrm{l}$ instead of $\boldsymbol{u}_\mathrm{h}$, using the same $\omega_\mathrm{u_l}$ as the DRL agents in Section~\ref{subsubsec:drl_mpc_case}. In its prediction, the vehicle splitting rate inputs are taken from the most recent high-level MPC trajectory, shifted by one step with the terminal value duplicated.

Both MPC controllers enforce the queue length limits as hard constraints. The two MPC problems are solved using the Matlab \texttt{fmincon} function with the SQP algorithm and the same tolerances as in Section~\ref{subsubsec:drl_mpc_case}. The high-level and low-level MPC controllers use a multi-start strategy with $5$ and $20$ initializations, respectively, tuned to balance the trade-off between computation time and optimality.

\subsubsection{State-feedback-MPC framework}\label{subsubsec:pi_alinea_mpc_case}
The state-feedback-MPC (SF-MPC) controller consists of a high-level MPC controller that computes the vehicle splitting rates and a low-level state-feedback controller, PI-ALINEA \citep{wang2006local}, that computes the ramp metering rates.

PI-ALINEA is a widely used state-feedback ramp metering method that updates the ramp metering rate using proportional and integral feedback of the measured bottleneck density. In particular, for each on-ramp, PI-ALINEA updates the ramp metering rate as
\begin{equation*}
u_{\mathrm{l}}(k_{\mathrm{l}}+1) = u_{\mathrm{l}}(k_{\mathrm{l}})
+ K_{\mathrm{R}}\big(\bar{\rho}-\rho_{\mathrm{b}}(k_{\mathrm{l}})\big)
- K_{\mathrm{A}}\big(\rho_{\mathrm{b}}(k_{\mathrm{l}})-\rho_{\mathrm{b}}(k_{\mathrm{l}}-1)\big),
\end{equation*}
where $u_{\mathrm{l}}$ denotes the ramp metering rate, $\rho_{\mathrm{b}}(k_{\mathrm{l}})$ is the lane-averaged density measured at the bottleneck location (i.e., the link directly downstream of the on-ramp), whereas the desired operating density $\bar{\rho}$ and the gains $K_{\mathrm{R}}$ and $K_{\mathrm{A}}$ are the PI-ALINEA parameters.

The high-level MPC controller uses the same formulation and parameter settings as in Section~\ref{subsubsec:drl_mpc_case}. Predicted ramp metering inputs required by the high-level MPC prediction are generated by running the PI-ALINEA controller inside the prediction model, similar to running the DRL policy within the high-level MPC prediction in the proposed DRL-MPC framework.

Separate PI-ALINEA parameters are tuned for the two on-ramps using the Bayesian optimization function of the Matlab Statistics and Machine Learning Toolbox. For each scenario defined in Section~\ref{subsec:case_study_setup}, Bayesian optimization is run five times independently, yielding five tuned parameter sets (see \ref{A2:PIalinea_params}), for a fair comparison with the five independent DRL training runs per scenario. Each run uses the same DelftBlue computing setting as for DRL training, with parallel evaluations enabled and a total runtime matched to the average DRL training time. Based on preliminary trial runs, the search ranges of the PI-ALINEA parameters are restricted to effective regions for the case study. Specifically, for each on-ramp $r\in\{1,2\}$, the parameter ranges are set as $K_{\mathrm{R},r}\in[0,4]$, $K_{\mathrm{A},r}\in[0,15]$, and $\bar{\rho}_{r}\in[30,100]$. The Bayesian optimization objective is computed by running the state-feedback-MPC controller over the simulation period and evaluating the soft objective cost (SOC) defined in \eqref{eq:eval_objective} below, with the only difference that the control input change penalty considers only the ramp metering inputs and not the vehicle splitting rate, since the vehicle splitting rate is included in the high-level MPC cost function and its variation is minimized by the high-level MPC controller.

\subsubsection{No-control case}\label{subsubsec:no_control_case}
The no-control case applies a constant vehicle splitting rate of $0.5$ and sets both ramp metering rates to $1$, i.e., vehicles are evenly split between the two downstream routes and on-ramp flows are not restricted. Note that this strategy yields reasonable performance in the considered case study due to the symmetric network structure (see Figure~\ref{fig:network_layout}) and the similar demand profiles at the two on-ramps (see Figure~\ref{fig:demands}).

\subsubsection{Evaluation metrics}\label{subsubsec:case_study_metrics}

The compared control frameworks are evaluated using the TTS, the queue length constraint violations, the total input variation (TIV), the SOC, and the total computation time as metrics.

Traffic efficiency is assessed by the TTS, obtained by accumulating the per-step cost $J_\mathrm{TTS}(\boldsymbol{x}(k))$ over the simulation period. The per-step TTS cost is defined as
\begin{equation}\label{eq:JTTS_case}
J_\mathrm{TTS}(\boldsymbol{x}(k))
= T \sum_{c=1}^{C}\left(
\sum_{m\in\mathcal{M}} \sum_{i\in\mathcal{I}_m} \rho_{m,i,c}(k)\, L_m\, \lambda_m
+\sum_{o\in\{O_1,O_2,O_3\}} w_{o,c}(k)
\right), 
\end{equation}
where $\mathcal{M}$ is the set of links, $\mathcal{I}_m$ is the set of segments in link $m$, $\rho_{m,i,c}(k) \in \boldsymbol{x}(k)$ is the density of class $c$ in segment $i$ of link $m$ at network sampling step $k$, $w_{o,c}(k) \in \boldsymbol{x}(k)$ is the queue length of class $c$ at origin $o$ at network sampling step $k$, and $L_m$ and $\lambda_m$ denote the segment length and the number of lanes in link $m$, respectively (see \ref{A1:Metanet} for details).

Constraint handling is assessed using the queue length limit exceedance at origin $o \in \{O_1, O_2, O_3\}$ at network sampling step $k$, defined as
\begin{equation}\label{eq:queue_exceed}
\Delta w_{o}(k) = \max\!\left(0,\, w_o(k) - w_o^{\max}\right).
\end{equation}
The total queue length constraint violation over the simulation period is defined as
\begin{equation}\label{eq:queue_exceed_total}
\Delta w_{\mathrm{tot}} = \sum_{o \in \{O_1, O_2, O_3\}} \sum_{k \in \mathcal{K}} \Delta w_{o}(k),
\end{equation}
where $\mathcal{K}$ denotes the set of simulation time step indices. Thus, $\Delta w_{\mathrm{tot}}$ is reported with number of vehicles as units. The maximum queue length constraint violation over the simulation period is defined as
\begin{equation}\label{eq:queue_exceed_max}
\Delta w_{\max} = \max_{o \in \{O_1, O_2, O_3\},\, k \in \mathcal{K}}\, \Delta w_{o}(k).
\end{equation}

The total input variation accumulates the magnitude of the changes in the applied control inputs over the simulation period and is defined as
\begin{equation}\label{eq:tv_metric}
\mathrm{TIV} = \sum_{k \in \mathcal{K}\setminus\{0\}}
\lVert \boldsymbol{u}(k)-\boldsymbol{u}(k-1)\rVert_2,
\end{equation}
where $\boldsymbol{u}(k)$ stacks the vehicle splitting rate and the two ramp metering rates applied to the network at network sampling step $k$. A larger TIV indicates more aggressive variations in the vehicle splitting rate or in the ramp metering rates across consecutive control steps.

The SOC aggregates TTS, a quadratic penalty on consecutive input differences, and the soft quadratic queue-length-violation penalty into a single composite metric:
\begin{equation}\label{eq:eval_objective}
\begin{aligned}
\mathrm{SOC} ={}& \mathrm{TTS}
 + \omega_u \sum_{k \in \mathcal{K}\setminus\{0\}}
\lVert \boldsymbol{u}(k)-\boldsymbol{u}(k-1)\rVert_2^2
\\
& + \sum_{o \in \{O_1,O_2,O_3\}}\sum_{k \in \mathcal{K}} \Delta w_o(k)^2,
\end{aligned}
\end{equation}
where the value of $\omega_u$ is set as $\omega_\mathrm{u_l}/m_\mathrm{l}$ since the input trajectories $\boldsymbol{u}(k)$ are evaluated at the network sampling time step rather than at the low-level control step.

Computational efficiency is assessed by the total computation time required to compute the control inputs over the simulation period.

\subsection{Training results}\label{subsec:training_results}
\begin{figure}[tbp]
    \centering
    \includegraphics[width=0.75\textwidth]{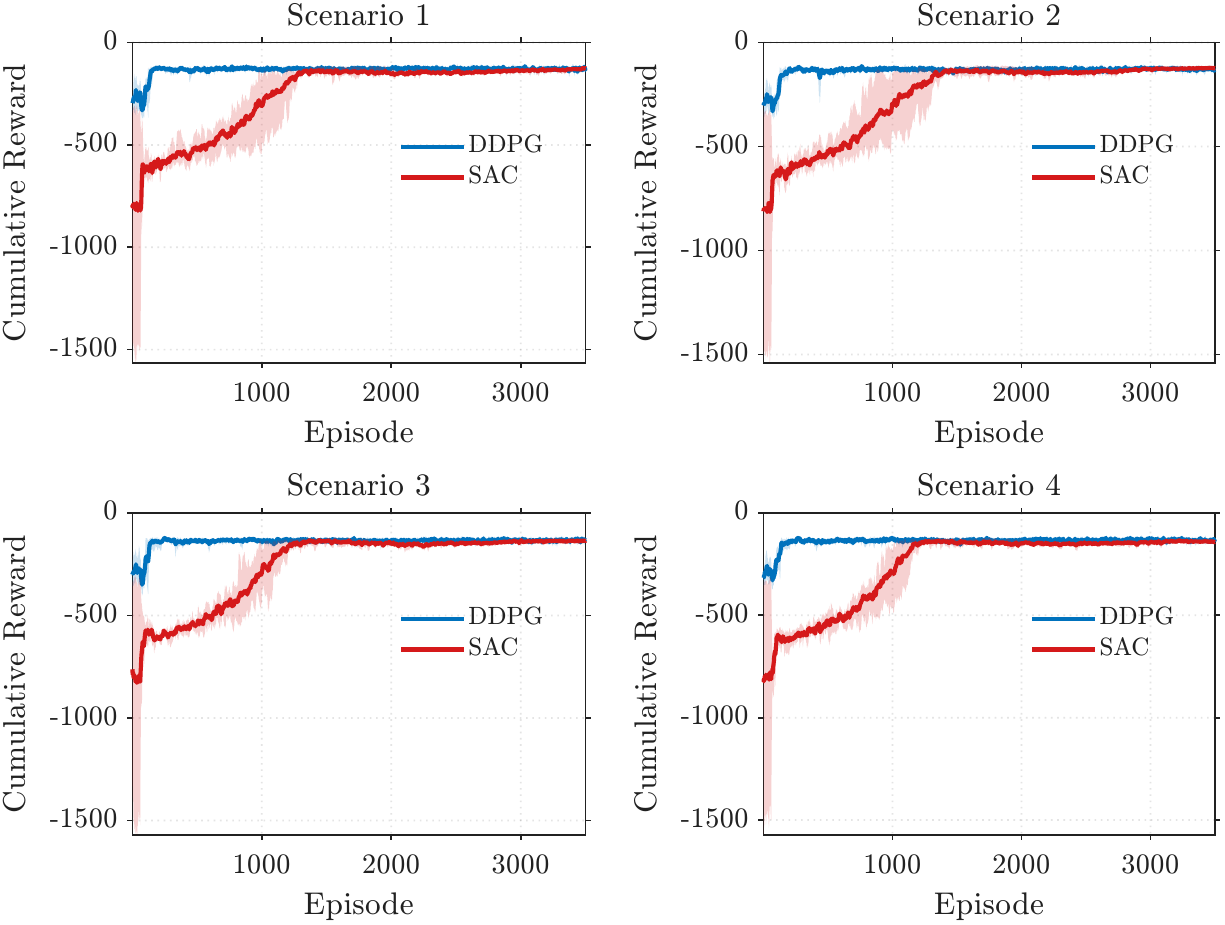}
    \caption{Cumulative reward curves of the five DDPG and five SAC agents under four scenarios. Bold curves show the mean cumulative reward across the five agents and shaded areas show the min--max range.}
    \label{fig:learning_curves}
\end{figure}

Figure~\ref{fig:learning_curves} reports the learning curves of the five DDPG and five SAC agents, trained with independent random seeds in each of the four scenarios. The plotted returns retain the $1/30$ reward scaling used during training and are smoothed using a 10-episode moving average before computing the across-seed statistics, making the learning trends easier to discern. Each curve shows the per-episode return, defined as the undiscounted sum of rewards collected within an episode.

In all four scenarios, the learning curves of all agents converge within the training budget, and the across-seed min--max ranges narrow as training progresses. Note that, since the plotted returns are obtained during training under exploration, they do not reflect the deployment performance of the converged policies. The deployment performance of the agents within the DRL-MPC framework is reported in the next section.

\subsection{Deployment results}\label{subsec:deployment_results}

The deployment results include two variants of the proposed DRL-MPC framework: DDPG-MPC, which deploys the deterministic DDPG actor, and SAC-MPC, which deploys the trained SAC actor deterministically by computing the ramp metering rates from the mean of its output distribution, following the observation of \citet{haarnoja2018soft} that deterministic evaluation of a trained SAC policy often achieves better performance. Per scenario, the hierarchical MPC framework and the no-control case are assessed over the $10$ evaluation simulations, each using a different random seed, while SF-MPC and each DRL-MPC variant are assessed over $50$ runs, obtained by repeating these same $10$ evaluation simulations for the five tuned PI-ALINEA parameter sets and the five independently trained DRL agents, respectively.

\begin{figure}[tbp]
    \centering
    \includegraphics[width=0.75\textwidth]{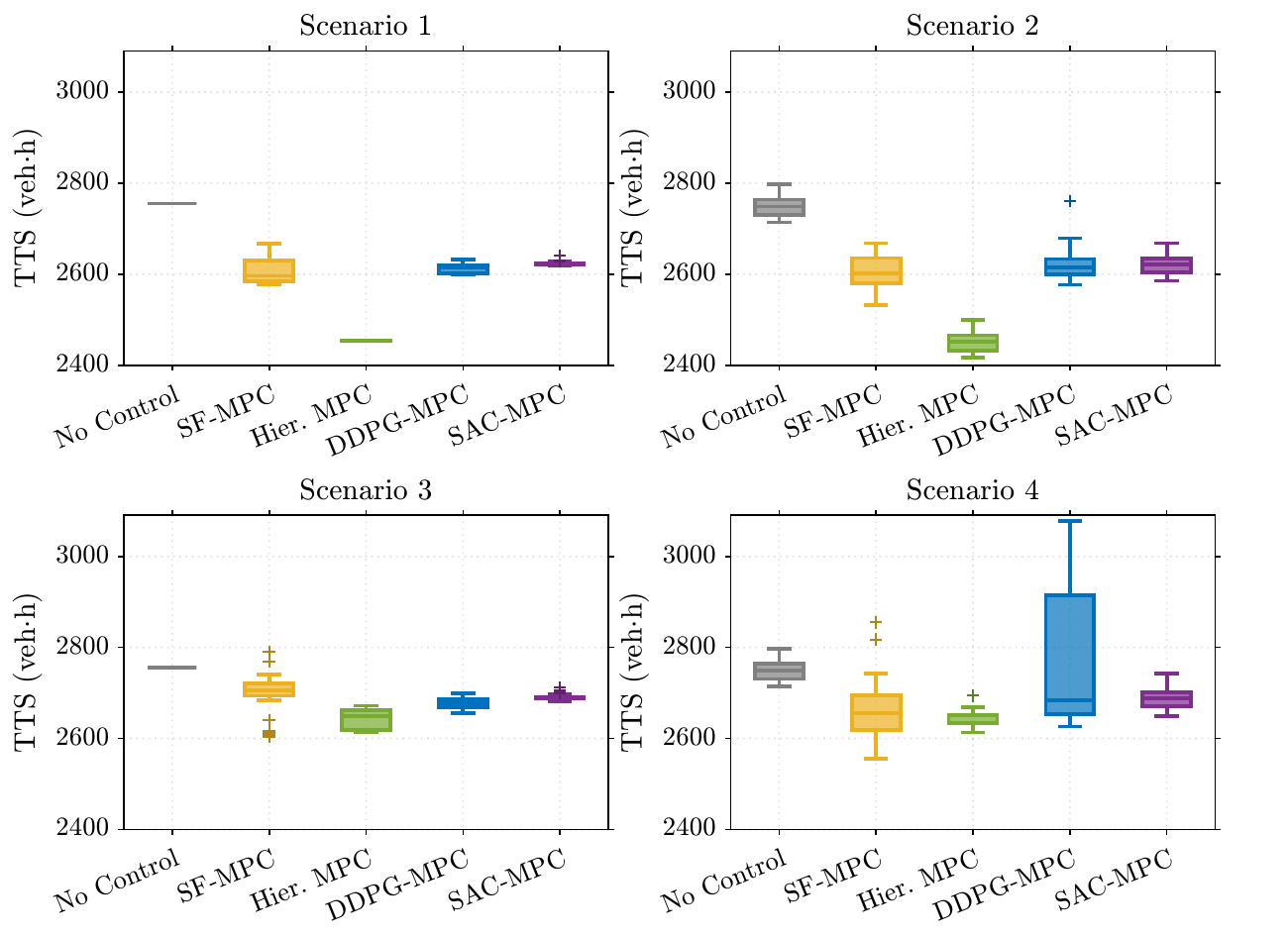}
    \caption{Total Time Spent (TTS) comparison for all control frameworks under the four evaluation scenarios.}
    \label{fig:tts_comparison}
\end{figure}

Figure~\ref{fig:tts_comparison} reports the TTS obtained by each control framework in the four scenarios. In Scenario~1 and Scenario~2, where the MPC-based controllers have access to a perfect model of the network dynamics, the hierarchical MPC framework achieves the lowest median TTS, with a reduction of about $10\%$ relative to the no-control case. DDPG-MPC, SAC-MPC, and SF-MPC obtain similar median TTS values in these two scenarios, with reductions of about $5\%$ relative to the no-control case.

\begin{figure}[tbp]
    \centering
    \includegraphics[width=0.75\textwidth]{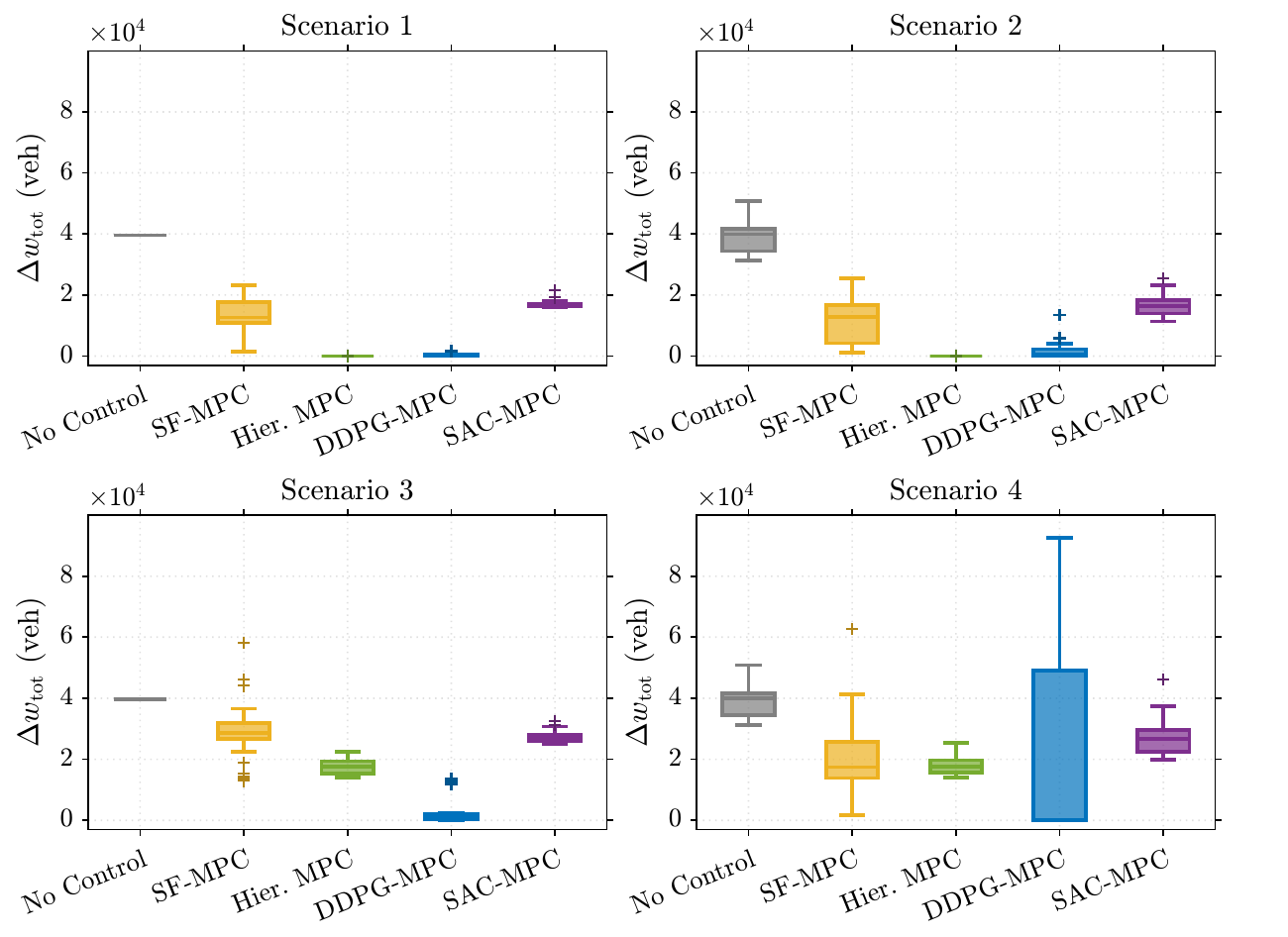}
    \caption{Total queue length constraint violation comparison for all control frameworks under the four evaluation scenarios.}
    \label{fig:dw_tot_comparison}
\end{figure}

\begin{figure}[tbp]
    \centering
    \includegraphics[width=0.75\textwidth]{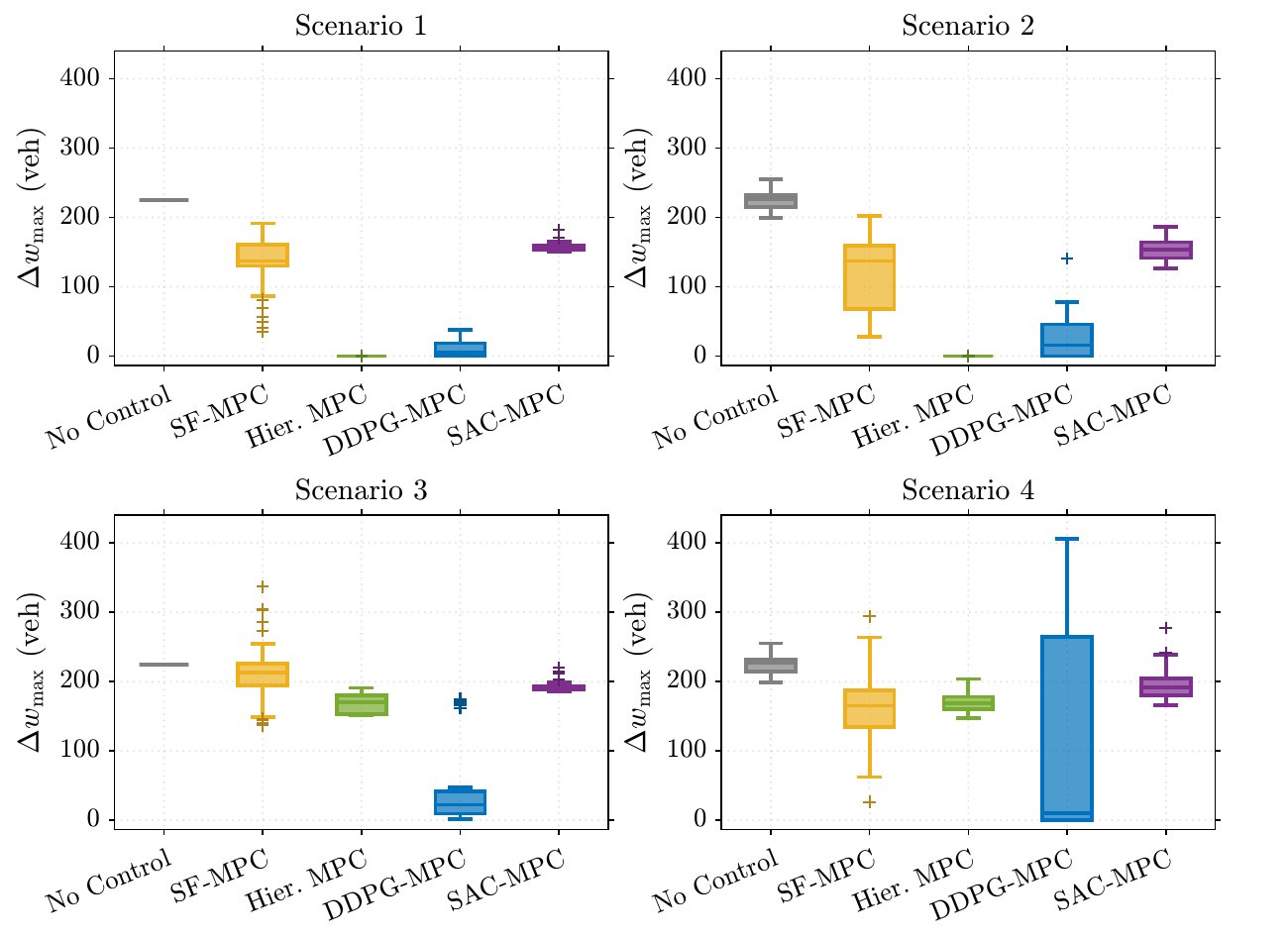}
    \caption{Maximum queue length constraint violation comparison for all control frameworks under the four evaluation scenarios.}
    \label{fig:dw_max_comparison}
\end{figure}

\begin{figure}[tbp]
    \centering
    \includegraphics[width=0.75\textwidth]{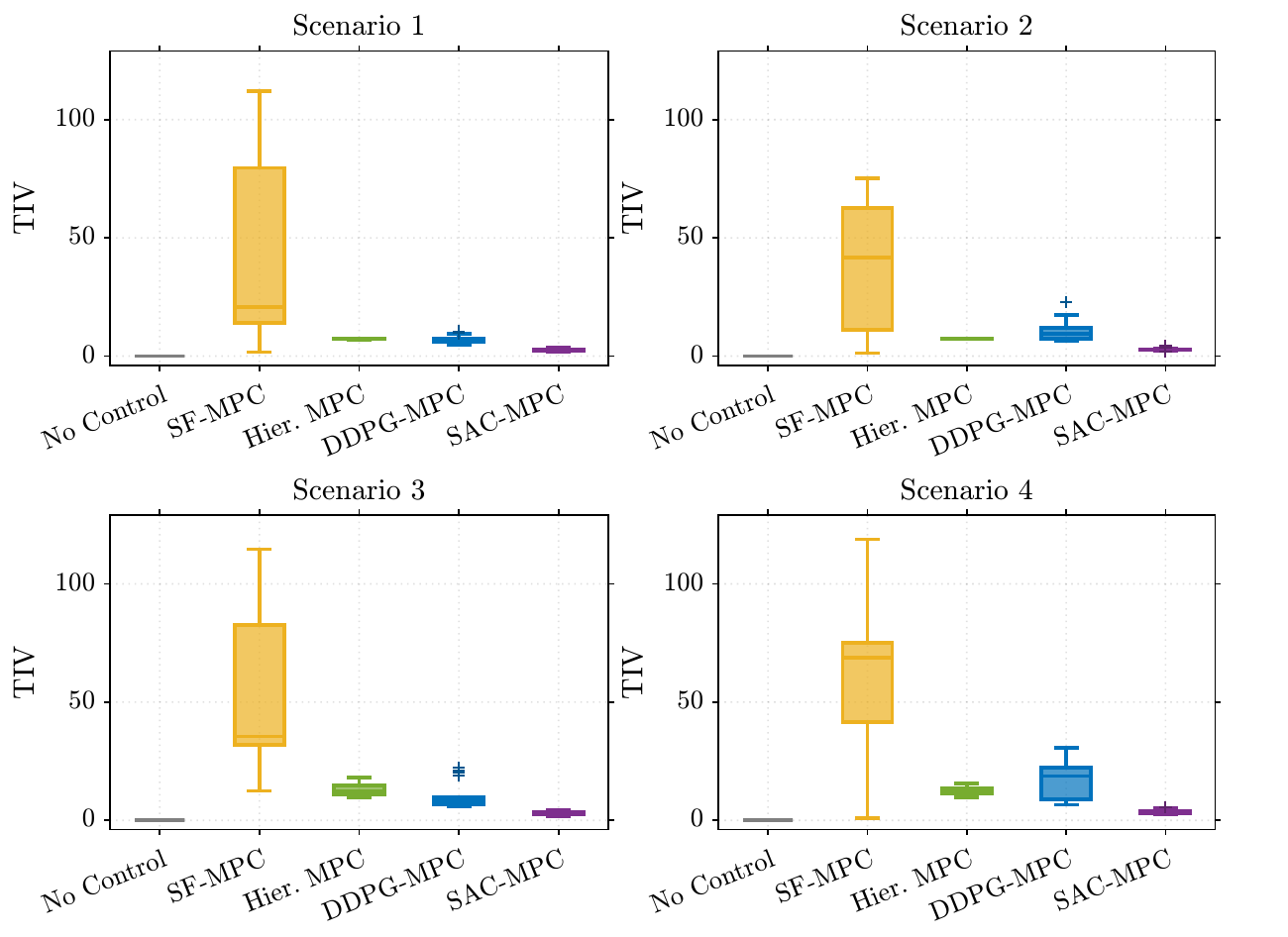}
    \caption{Total input variation (TIV) comparison for all control frameworks under the four evaluation scenarios.}
    \label{fig:tv_comparison}
\end{figure}

\begin{figure}[tbp]
    \centering
    \includegraphics[width=0.75\textwidth]{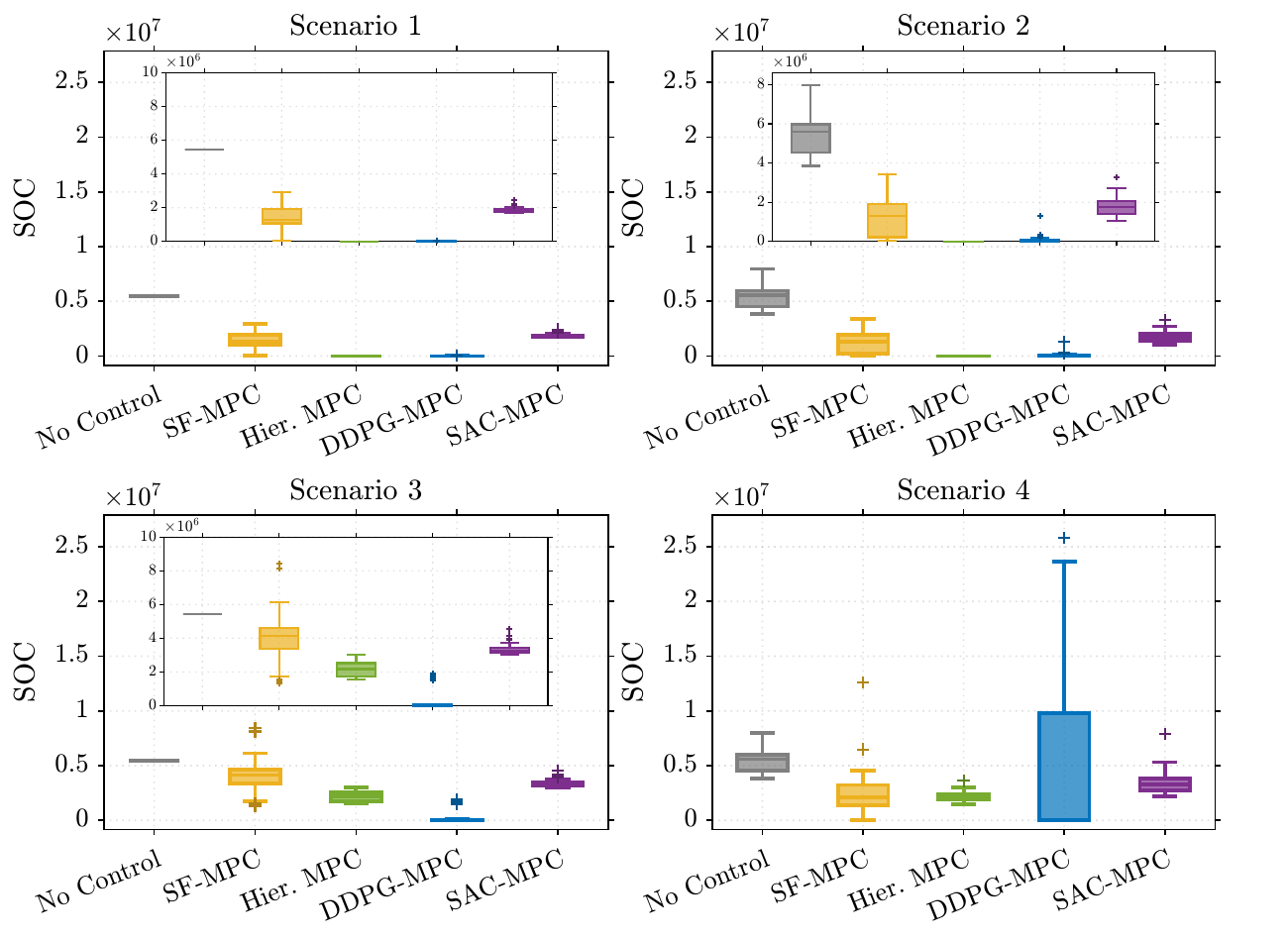}
    \caption{Soft objective cost (SOC) comparison, where SOC combines TTS, input-variation penalty, and the soft quadratic queue-length-violation penalty.}
    \label{fig:obj_comparison}
\end{figure}

Under prediction model mismatch (Scenario~3 and Scenario~4), the median TTS advantage of the hierarchical MPC framework shrinks to about $4\%$ relative to the no-control case, less than half of the reduction achieved when the prediction model is matched. DDPG-MPC and SAC-MPC remain competitive with the hierarchical MPC framework in terms of median TTS in both mismatch scenarios, with SAC-MPC trailing DDPG-MPC by a small margin. The TTS spread of DDPG-MPC widens noticeably in Scenario~4, where one of the five trained agents attains substantially worse performance than the remaining four, which inflates the across-seed range. This behavior is also observed in the remaining performance metrics (see Figures~\ref{fig:dw_tot_comparison}--\ref{fig:obj_comparison}) and is consistent with the higher variability across random training seeds reported for DDPG relative to SAC in~\citep{haarnoja2018soft}.

Figures~\ref{fig:dw_tot_comparison} and~\ref{fig:dw_max_comparison} report the total and the maximum queue length constraint violations, respectively. In Scenario~1 and Scenario~2, the hierarchical MPC framework keeps the queue length constraint violations close to zero, since it explicitly accounts for the queue length limits during optimization using a prediction model that matches the simulated network. DDPG-MPC also keeps queue length limit violations small: the total queue length constraint violation is close to zero in Scenario~1 and remains low in Scenario~2, while the median maximum queue length constraint violation is substantially smaller than for SF-MPC, SAC-MPC, and the no-control case. When the prediction model is accurate, DDPG-MPC closely matches the constraint-handling performance of the hierarchical MPC framework.

Under prediction model mismatch (Scenario~3 and Scenario~4), the differences between the control frameworks become more pronounced. The hierarchical MPC framework can no longer keep the queue length constraint violations close to zero in the simulated network, since its constraint handling operates on a mismatched prediction model. SF-MPC also incurs larger median total and maximum queue length constraint violations. DDPG-MPC, in contrast, keeps the median total and maximum queue length constraint violations small, showing that the learned low-level ramp metering policy provides useful robustness against errors in the high-level MPC prediction model. SAC-MPC, however, incurs larger median total and maximum queue length constraint violations than DDPG-MPC in both mismatch scenarios.

Figure~\ref{fig:tv_comparison} reports the TIV values.

SF-MPC generates large TIV values in most scenarios, consistent with the known tendency of PI-ALINEA-based state-feedback ramp metering controllers to produce substantial variations in the ramp metering rates~\citep{airaldi2025reinforcement}. The hierarchical MPC framework and DDPG-MPC produce smoother control trajectories, with DDPG-MPC remaining close to the hierarchical MPC framework in terms of median TIV in all scenarios. SAC-MPC attains the smallest TIV among the controlled methods, which may be attributed to the entropy regularization employed during SAC training. By promoting stochastic exploration throughout training, entropy regularization may yield learned mean actions that vary less aggressively between consecutive control steps when deployed deterministically.

Figure~\ref{fig:obj_comparison} reports the SOC values.

In Scenario~1 and Scenario~2, the hierarchical MPC framework attains the lowest median SOC by minimizing the TTS while keeping the queue length constraint violations close to zero and the input variations small, owing to a prediction model that matches the simulated network. DDPG-MPC obtains a lower median SOC than SAC-MPC and SF-MPC in these two scenarios. The higher SOC values of SAC-MPC and SF-MPC are mostly driven by their larger queue length constraint violations, owing to the quadratic queue-length-violation penalty in the SOC. Under prediction model mismatch, DDPG-MPC achieves the lowest median SOC among all controlled methods in both Scenario~3 and Scenario~4, whereas SAC-MPC attains higher median values. This indicates that the learned DDPG ramp metering policy compensates more effectively for mismatch in the high-level MPC prediction model.

Table~\ref{tab:comp_time_comparison} reports the mean and standard deviation of the total control framework computation time over the simulation period across all experimental runs.

\begin{table}[tbp]
\centering
\caption{Mean and standard deviation of the total control framework computation time (s) across all experimental runs for each control framework and scenario.}
\label{tab:comp_time_comparison}
\small
\begin{tabular}{l r@{$\,\pm\,$}l r@{$\,\pm\,$}l r@{$\,\pm\,$}l r@{$\,\pm\,$}l}
\toprule
\textbf{Controller} & \multicolumn{2}{c}{\textbf{Scenario 1}} & \multicolumn{2}{c}{\textbf{Scenario 2}} & \multicolumn{2}{c}{\textbf{Scenario 3}} & \multicolumn{2}{c}{\textbf{Scenario 4}} \\
\midrule
SF-MPC & $30.5$ & $1.9$ & $29.0$ & $2.2$ & $27.7$ & $1.0$ & $27.9$ & $1.2$ \\
Hier.~MPC & $1778.9$ & $3.7$ & $1736.7$ & $5.1$ & $1638.2$ & $33.0$ & $1659.3$ & $39.2$ \\
DDPG-MPC & $55.8$ & $3.5$ & $52.9$ & $2.3$ & $54.6$ & $2.8$ & $54.3$ & $2.3$ \\
SAC-MPC & $76.9$ & $2.6$ & $78.8$ & $3.1$ & $62.4$ & $1.8$ & $64.9$ & $2.1$ \\
\bottomrule
\end{tabular}
\end{table}

Here, the no-control case is omitted, since it involves no control input computation. The hierarchical MPC framework requires by far the largest computation time, around $28$~minutes across the four scenarios, since it solves a nonlinear optimization problem at both the high-level and the low-level control time scales. The two DRL-MPC variants have computation times that are close to each other, in the order of a minute, which corresponds to approximately a thirtyfold reduction relative to the hierarchical MPC framework. SF-MPC is the fastest control framework, since the PI-ALINEA state-feedback controller requires less computation time than a forward pass through the DRL policy network.

Overall, the comparison shows that DDPG-MPC provides the most favorable balance among the proposed DRL-MPC variants and the considered baselines: it preserves a competitive median TTS across the four scenarios, substantially reduces the median queue length constraint violations under prediction model mismatch relative to SF-MPC and the hierarchical MPC framework, and reduces the online computation time by approximately a factor of thirty compared with the hierarchical MPC framework. Although SAC-MPC produces smoother control trajectories and shows lower variability across training seeds, it incurs larger queue length constraint violations and higher SOC values than DDPG-MPC. A possible explanation for this performance difference is the distinct policy-learning mechanisms of the two algorithms: DDPG learns a DRL policy using deterministic policy gradients, whereas SAC learns a DRL policy using an entropy-regularized objective that balances expected return and policy entropy.

\clearpage

\section{Conclusions}\label{sec:conclusion}

This paper has proposed a novel DRL-MPC framework that divides the control inputs between DRL and MPC, with an application to the control of multi-class transportation networks. By assigning the low-frequency control inputs to MPC and the high-frequency control inputs to DRL, the proposed framework combines the built-in optimization and constraint-handling capabilities of MPC with the fast online computation and model independence of DRL.

The proposed framework has been evaluated on a benchmark multi-class freeway network with one vehicle splitting rate control input, computed by the high-level MPC controller, and two ramp metering rate control inputs, computed by the low-level DRL agent. Across the scenarios that combine prediction model mismatch and noisy traffic demands, the proposed framework has been shown to offer a favorable trade-off between traffic efficiency, constraint handling, input smoothness, and online computation time, and to be more resilient to prediction model mismatch than a hierarchical MPC framework and a state-feedback-MPC framework.

Beyond the considered multi-class transportation network control case study, the proposed DRL-MPC framework could also be adapted to other control problems involving multiple control time scales, such as robotic motion control \citep{rosolia2022unified}, where reference trajectories are updated at a low frequency while actuator commands are updated at a high frequency, and power-electronic converter control \citep{xue2024multirate}, where motor-current references are updated at a low frequency while semiconductor switching commands are applied at a high frequency. Future work will consider evaluating the proposed framework across different application areas. Another direction will be to extend the proposed framework to the control of large-scale transportation networks by combining distributed MPC with multi-agent DRL.

\appendix
\section{Multi-class METANET model}
\label{A1:Metanet}

The METANET model \citep{kotsialos1999integrated} is a macroscopic traffic flow model widely used for simulating freeway networks \citep{airaldi2025reinforcement,chanfreut2020coalitional}, including multi-class freeway networks \citep{liu2016model}, due to its good balance between computational efficiency and modeling accuracy. In the current work, we employ the multi-class METANET model both to simulate the real multi-class freeway network and as the prediction model within the MPC controller.

In the multi-class METANET model \citep{liu2016model}, a freeway network consists of links representing freeway stretches with approximately uniform characteristics, connected by nodes representing junctions, ramps, or locations of major geometric changes. Each segment is described by the traffic density $\rho_{m,i,c}(k)$, mean speed $v_{m,i,c}(k)$, and outflow $q_{m,i,c}(k)$ of each vehicle class. Here, $m$ is the link index, $i$ is the segment index, $c \in \{1,2,\ldots,C\}$ is the vehicle class index, $C$ is the number of vehicle classes, and $k$ is the network sampling step counter. The number of segments in link $m$ is denoted by $N_m$. The density $\rho_{m,i,c}(k)$ is calculated as
\begin{equation*}
\rho_{m,i,c}(k) = \frac{L^\mathrm{veh}_c}{L^\mathrm{veh}_1} \rho^\mathrm{act}_{m,i,c}(k),
\end{equation*}
where $\rho^\mathrm{act}_{m,i,c}(k)$ is the actual number of vehicles of class $c$ per unit length and per lane, and $L^\mathrm{veh}_c$ and $L^\mathrm{veh}_1$ represent the typical vehicle lengths of vehicle class $c$ and the base class (taken as class $1$), respectively.

The segment outflow for vehicle class $c$ is calculated as
\begin{equation*}
q_{m,i,c}(k) = \rho_{m,i,c}(k)v_{m,i,c}(k)\lambda_m,
\end{equation*}
where $\lambda_m$ is the number of lanes in link $m$. The segment density for vehicle class $c$ is calculated as
\begin{equation*}
\rho_{m,i,c}(k+1) = \rho_{m,i,c}(k) + \frac{T}{L_{m} \lambda_m}\Big(q_{m,i-1,c}(k)-q_{m,i,c}(k)\Big),
\end{equation*}
where $T$ is the network sampling interval and $L_{m}$ is the length of the segments in link $m$.

The fraction of the traffic volume of class $c$ in a segment is calculated as
\begin{equation*}
\xi_{m,i,c}(k) = \frac{\rho_{m,i,c}(k)}{\rho_{m,i}(k)},
\end{equation*}
where the total density of the segment is defined as
\begin{equation*}
\rho_{m,i}(k) = \sum_{c=1}^C \rho_{m,i,c}(k).
\end{equation*}

The segment mean speed is calculated as
\begin{equation}\label{eq:v_m_i}
\begin{split}
v_{m,i,c}(k+1) = &v_{m,i,c}(k) + \frac{T}{\tau_c}\big(\tilde{V}_c(\rho_{m,i}(k),\Xi_{m,i}(k))-v_{m,i,c}(k)\big) \\
&+ \frac{T}{L_{m}}v_{m,i,c}(k)\big(v_{m,i-1,c}(k)-v_{m,i,c}(k)\big) \\
&- \frac{T \eta_c}{L_{m} \tau_c} \frac{\rho_{m,i+1}(k)-\rho_{m,i}(k)}{\rho_{m,i}(k)+\kappa_c},
\end{split}
\end{equation}
where $\Xi_{m,i}(k) = (\xi_{m,i,1}(k), \xi_{m,i,2}(k), \dots, \xi_{m,i,C}(k))$ and $\tau_c$, $\eta_c$, and $\kappa_c$ are model parameters. The term $\tilde{V}_c(\rho_{m,i}(k),\Xi_{m,i}(k))$ represents the interpolated speed-density relationship of vehicle class $c$, and it is calculated as
\begin{equation*}
\tilde{V}_c(\rho_{m,i}(k),\Xi_{m,i}(k)) = \min \Big(V_c(\rho_{m,i}(k)), \sum_{\gamma=1}^{C} \xi_{m,i,\gamma}(k) V_{\gamma}(\rho_{m,i}(k)) \Big).
\end{equation*}
This interpolation relies on $V_c(\rho_{m,i}(k))$, which denotes the desired speed of the drivers of class $c$ for the given density and is calculated as
\begin{equation*}
V_c(\rho_{m,i}(k)) = v^\mathrm{free}_{m,c} \exp\left[-\frac{1}{a_{m,c}} \left( \frac{\rho_{m,i}(k)}{\rho^\mathrm{crit}_{m}} \right)^{a_{m,c}} \right],
\end{equation*}
where $v^\mathrm{free}_{m,c}, a_{m,c}$, and $\rho^\mathrm{crit}_{m}$ are model parameters.

Links are connected at nodes, where the flow of each entering link is distributed among the leaving links according to a per-class splitting rate. Let \mbox{$\mathcal{L}^{\mathrm{in}}_n$} and \mbox{$\mathcal{L}^{\mathrm{out}}_n$} denote the sets of links that enter and leave node \mbox{$n$}, respectively. The total flow of vehicles of class \mbox{$c$} that enters node \mbox{$n$} is calculated as
\begin{equation*}
    Q_{n,c}(k) = \sum_{\mu \in \mathcal{L}^{\mathrm{in}}_n} q_{\mu,N_\mu,c}(k),
\end{equation*}
and the inflow of vehicles of class \mbox{$c$} into the first segment of a leaving link \mbox{$m \in \mathcal{L}^{\mathrm{out}}_n$} is calculated as
\begin{equation*}
    q_{m,1,c}(k) = \beta_{n,m,c}(k)\, Q_{n,c}(k),
\end{equation*}
where \mbox{$\beta_{n,m,c}(k) \in [0,1]$} is the splitting rate of class \mbox{$c$} at node \mbox{$n$} towards leaving link \mbox{$m$}, with \mbox{$\sum_{m \in \mathcal{L}^{\mathrm{out}}_n} \beta_{n,m,c}(k) = 1$}. When node \mbox{$n$} has more than one leaving link, the virtual downstream density used in \mbox{\eqref{eq:v_m_i}} for the last segment of an entering link \mbox{$m \in \mathcal{L}^{\mathrm{in}}_n$} is calculated as
\begin{equation*}
    \rho_{m,N_m+1}(k) = \frac{\sum_{\mu \in \mathcal{L}^{\mathrm{out}}_n}\rho_{\mu,1}^{2}(k)}{\sum_{\mu \in \mathcal{L}^{\mathrm{out}}_n}\rho_{\mu,1}(k)},
\end{equation*}
where \mbox{$\rho_{\mu,1}(k)$} is the total density of the first segment of leaving link \mbox{$\mu$}.

The queue length at an origin (e.g., a mainstream source or an on-ramp) is updated as
\begin{equation*}
    w_{o,c}(k+1) = w_{o,c}(k)+T(d_{o,c}(k)-q_{o,c}(k)),
\end{equation*}
where $d_{o,c}(k)$ and $q_{o,c}(k)$ are the demand and the outflow of vehicles of class $c$ at origin $o$ at network sampling step $k$, respectively.

For an on-ramp, $q_{o,c}(k)$ is calculated as
\begin{equation*}
\begin{split}
    q_{o,c}(k) = \min \Bigg[ &q^\mathrm{des}_{o,c}(k), \frac{q^\mathrm{des}_{o,c}(k)}{\sum_{\gamma=1}^{C}q^\mathrm{des}_{o,\gamma}(k)} Q_o r_{o}(k), \\
    &\frac{q^\mathrm{des}_{o,c}(k)}{\sum_{\gamma=1}^{C}q^\mathrm{des}_{o,\gamma}(k)} Q_o \left( \frac{\rho^\mathrm{max}_{\mu} -\rho_{\mu,1}(k)}{\rho^\mathrm{max}_{\mu} - \rho^\mathrm{crit}_{\mu}} \right) \Bigg],
\end{split}
\end{equation*}
where $\mu$ is the index of the link to which the on-ramp is connected, {$Q_o$} is the free-flow capacity of on-ramp origin, $r_o(k) \in [0,1]$ is the ramp metering rate at network sampling step $k$, and $\rho^\mathrm{max}_{\mu}$ and $\rho^\mathrm{crit}_{\mu}$ are the maximum and critical densities of link $\mu$, respectively. The term $q^\mathrm{des}_{o,c}(k)$ denotes the desired origin outflow for class $c$ at origin $o$ and is calculated as
\begin{equation*}
    q^\mathrm{des}_{o,c}(k) = d_{o,c}(k) + \frac{w_{o,c}(k)}{T}.
\end{equation*}

For mainstream sources, $q_{o,c}(k)$ is calculated as
\begin{equation*}
\displaystyle\begin{aligned}
    q_{o,c}(k) = \min \Bigg[ &q^\mathrm{des}_{o,c}(k), \frac{q^\mathrm{des}_{o,c}(k)}{\sum_{\gamma=1}^{C}q^\mathrm{des}_{o,\gamma}(k)} Q^\mathrm{main}_o, \\
    &\frac{q^\mathrm{des}_{o,c}(k)}{\sum_{\gamma=1}^{C}q^\mathrm{des}_{o,\gamma}(k)} Q^\mathrm{main}_o \left( \frac{\rho^\mathrm{max}_{\mu} -\rho_{\mu,1}(k)}{\rho^\mathrm{max}_{\mu} - \rho^\mathrm{crit}_{\mu}} \right) \Bigg],
\end{aligned}
\end{equation*}
where $Q^\mathrm{main}_o$ is the free-flow capacity of mainstream origin aggregated over the lanes of the link it feeds.

For links with an on-ramp, the mean speed of the first segment is adjusted to account for the speed drop caused by merging phenomena by adding the term
\begin{equation*}
- \frac{T \sigma_c q_o(k) v_{m,1,c}(k)}{L_{m} \lambda_m \left(\rho_{m,1}(k) + \kappa_c \right)}
\end{equation*}
to \eqref{eq:v_m_i}, where $q_{o}(k) = \sum_{\gamma=1}^{C} q_{o,\gamma}(k)$ is the total outflow from the on-ramp and $\sigma_c$ is a model parameter.

\section{Training algorithms}
\label{app:training}

This appendix details the two actor-critic algorithms used to train the DRL agent within the DRL-MPC framework, namely the Deep Deterministic Policy Gradient (DDPG) and the Soft Actor-Critic (SAC) algorithms, together with the overall training procedure and the network architectures and hyperparameters used for the two algorithms in the case study of Section~\ref{sec:case_study}.

\subsection{Training the framework using DDPG}
\label{subsec:train_ddpg}

The DDPG algorithm \citep{lillicrap2015continuous} is an off-policy actor-critic approach designed for environments with continuous action spaces. It learns a deterministic policy through an actor network $\pi_\theta$, parameterized by $\theta$, which maps the observed state to a DRL action, while employing a critic network $Q_\phi$, parameterized by $\phi$, which estimates the expected discounted return for a state--action pair under the current policy (see Figure~\ref{RL_MPC_fig}). To enhance learning stability, DDPG also maintains a target actor network $\pi_{\theta'}$ and a target critic network $Q_{\phi'}$, whose parameters are updated gradually, ensuring smoother changes compared to the main actor and critic networks. All transitions $(\boldsymbol{x}_\mathrm{rl},\boldsymbol{u}_\mathrm{l},r,\boldsymbol{x}^{+}_\mathrm{rl})$ are collected from the environment and stored in a replay buffer, removing the oldest transitions once the buffer reaches its capacity. The actor and the critic are updated using a mini-batch of $N$ transitions sampled uniformly at random from the replay buffer. Sampling from the replay buffer reduces temporal correlation between training samples and enables data reuse, thereby improving learning stability and sample efficiency. For more details, the interested reader is referred to \citep{lillicrap2015continuous}.

\Citet{sun2024novel} showed that using the $n$-step temporal difference (TD) return, which sums the rewards over the next $n$ steps using a set of $n$ consecutive transitions sampled from the replay buffer, instead of the traditional one-step TD target improves the learning efficiency and enhances the quality of actions taken in systems with time delays, such as transportation networks. Therefore, we use the $n$-step TD method to compute the target value, which serves as the learning reference by representing the estimated total discounted reward, as
\begin{equation*}
    y({k_\mathrm{l}}) = r_n(k_\mathrm{l}) + \gamma^n Q_{\phi'}\big(\boldsymbol{x}_\mathrm{rl}(k_\mathrm{l}+n), \boldsymbol{u}^{\theta'}_\mathrm{l}(k_\mathrm{l}+n)\big),
\end{equation*}
where $\boldsymbol{u}^{\theta'}_\mathrm{l}(k_\mathrm{l}+n) = \pi_{\theta'}(\boldsymbol{x}_{\mathrm{rl}}(k_\mathrm{l}+n))$ and the $n$-step reward is calculated as
\begin{equation}\label{eq:n_step_reward}
    r_n(k_\mathrm{l}) = \sum^{n-1}_{o=0} \gamma^o r (k_\mathrm{l}+o).
\end{equation}
If an episode terminates within the $n$-step window, only the rewards collected up to termination are included in the target.

The loss for the critic network is calculated as
\begin{equation}\label{eq:ddpg_loss}
    L(\phi) = \frac{1}{N} \sum^{N}_{i=1} \Big(y(i) - Q_\phi (\boldsymbol{x}_\mathrm{rl}(i),\boldsymbol{u}_\mathrm{l}(i))\Big)^2,
\end{equation}
where the index $i$ runs over the data points of a mini batch of size $N$ that is randomly sampled from the replay buffer.

The parameters of the critic, $\phi$, are updated by minimizing the loss function \eqref{eq:ddpg_loss}, e.g., using the adaptive moment estimation (Adam) optimizer \citep{kingma2017adammethodstochasticoptimization}. Then, the parameters of the actor, $\theta$, are subsequently updated by maximizing the estimated return, $Q_\phi$, based on the policy gradient \citep{lillicrap2015continuous}.

The parameters of the target actor, $\theta'$, and the target critic, $\phi'$, are updated using Polyak averaging as
\begin{equation*}
\begin{aligned}
    \theta' \leftarrow \tau \theta + (1-\tau)\theta', \\
    \phi' \leftarrow \tau \phi + (1-\tau)\phi',
\end{aligned}
\end{equation*}
where $\tau$ is the smoothing factor, with $\tau \ll 1$, so that the target values change slowly, improving learning stability \citep{lillicrap2015continuous}.

To enhance exploration efficiency, noise is added to the DRL actions, e.g., using the Ornstein-Uhlenbeck process \citep{uhlenbeck1930theory}, which generates temporally correlated noise and can be particularly beneficial in control settings with inertia-like dynamics, such as those encountered in transportation network control.

\begin{algorithm}
\caption{Training algorithm for the DRL-MPC framework}
\label{alg:training}
\begin{algorithmic}[1]
\fontsize{10}{11}\selectfont
\renewcommand{\algorithmicrequire}{\textbf{Input:}}
\renewcommand{\algorithmicensure}{\textbf{Output:}}

\REQUIRE Initial neural-network parameters ($\theta, \theta',\phi$, and $\phi'$ for DDPG; \\ $\theta,\phi_1,\phi_2,\phi'_1$, and $\phi'_2$ for SAC), $m_\mathrm{h}$, $m_\mathrm{l}$, $\boldsymbol{\hat{d}}$, $k^{\text{max}}_{\text{eps}}$, $M$, $N$, $n$, $N_\mathrm{w}$, $N_\mathrm{e}$, $N_\mathrm{mb}^{\max}$, $N_\mathrm{t}$, and $\tau$
\STATE Initialize critic network(s) and actor network
\STATE Initialize target networks
\STATE Initialize experience replay buffer $\mathcal{D}$
\STATE Initialize critic-update counter $k_\phi \leftarrow 0$
\FOR{episode $= 1$ to $M$}
    \STATE Receive initial transportation network state $\boldsymbol{x}(0)$
    \STATE Initialize step counter $k \leftarrow 0$
    \STATE Observe initial DRL state $\boldsymbol{x}_\mathrm{rl}$

    \WHILE{$k < k^{\text{max}}_{\text{eps}}$}
        
        \IF{$k \pmod{m_\mathrm{h}} = 0$}
            \STATE $k_\mathrm{h} \leftarrow k/m_\mathrm{h}$
            \STATE Obtain $\boldsymbol{u}_\mathrm{h}(k_\mathrm{h})$ by solving \eqref{eq:MPC}
        \ENDIF
        \STATE $\boldsymbol{\bar{u}}_\mathrm{h}(k) \leftarrow \boldsymbol{u}_\mathrm{h}(k_\mathrm{h})$
        
        \IF{$k \pmod{m_\mathrm{l}} = 0$}
            \STATE $k_\mathrm{l} \leftarrow k/m_\mathrm{l}$
            \STATE Sample $\boldsymbol{u}_\mathrm{l}(k_\mathrm{l}) \sim \pi_\theta(\cdot \mid \boldsymbol{x}_\mathrm{rl}(k_\mathrm{l}))$; for DDPG, add exploration noise
        \ENDIF
        \STATE $\boldsymbol{\bar{u}}_\mathrm{l}(k) \leftarrow \boldsymbol{u}_\mathrm{l}(k_\mathrm{l})$
        
        \STATE $\boldsymbol{x}(k+1) \leftarrow F(\boldsymbol{x}(k), \boldsymbol{\bar{u}}_\mathrm{h}(k), \boldsymbol{\bar{u}}_\mathrm{l}(k), \boldsymbol{d}(k))$

        \IF{$(k+1) \pmod{m_\mathrm{l}} = 0$}
            \STATE Calculate reward $r(k_\mathrm{l})$ using \eqref{eq:RLreward}
            \STATE Observe new DRL state $\boldsymbol{x}_\mathrm{rl}(k_\mathrm{l}+1)$
            \STATE Store transition $(\boldsymbol{x}_\mathrm{rl}(k_\mathrm{l}), \boldsymbol{u}_\mathrm{l}(k_\mathrm{l}), r(k_\mathrm{l}), \boldsymbol{x}_\mathrm{rl}(k_\mathrm{l}+1))$ in $\mathcal{D}$
        \ENDIF
        
        \STATE $k \leftarrow k + 1$
        
    \ENDWHILE

    \IF{$|\mathcal{D}| \geq N_\mathrm{w}$}
        \STATE $N_\mathrm{mb} \leftarrow \min\!\left\{N_\mathrm{mb}^{\max},\left\lfloor |\mathcal{D}|/N \right\rfloor\right\}$
        \FOR{each of $N_\mathrm{mb}$ mini-batches in each of $N_\mathrm{e}$ epochs}
            \STATE Sample $N$ transitions from $\mathcal{D}$, each containing $n$ steps
            \STATE Update critic(s) by minimizing loss $L(\phi)$
            \STATE $k_\phi \leftarrow k_\phi+1$
            \STATE Update actor by maximizing the policy objective
            \STATE Update the entropy weight for SAC
            \IF{$k_\phi \pmod{N_\mathrm{t}}=0$}
                \STATE Update the target critic(s) and the target actor using Polyak averaging
            \ENDIF
        \ENDFOR
    \ENDIF
    
\ENDFOR
\ENSURE Trained actor policy $\pi_\theta$
\end{algorithmic}
\end{algorithm}

\subsection{Training the framework using SAC}
\label{subsec:train_sac}
The SAC algorithm \citep{haarnoja2018soft} is an off-policy actor-critic approach that can handle continuous action spaces. SAC learns a stochastic policy that optimizes a combination of the expected return and the policy entropy. The policy entropy quantifies the uncertainty of the policy given a particular state, with higher entropy encouraging broader exploration of the action space. By jointly maximizing the expected cumulative discounted reward and the entropy, SAC aims to achieve a balance between exploration and exploitation. SAC employs two critic networks and uses the minimum of their $Q$-value estimates to reduce positive overestimation bias. SAC utilizes a replay buffer and updates the actor and critics using mini-batches, as in DDPG. While SAC uses target critic networks in a manner similar to DDPG, it does not maintain a separate target actor network. Instead, the target critics evaluate actions generated directly by the actor $\pi_\theta$. For more details, the interested reader is referred to \citep{haarnoja2018soft}.

Similar to DDPG, the target value is computed via the $n$-step TD method using \eqref{eq:n_step_reward} as follows:
\begin{equation*}
\begin{aligned}
    y({k_\mathrm{l}}) = r_n(k_\mathrm{l}) + \gamma^n \Big(
    \min_{\psi \in \{1,2\}} Q_{\phi'_\psi}\big(\boldsymbol{x}_\mathrm{rl}(k_\mathrm{l}+n), \boldsymbol{u}_\mathrm{l}(k_\mathrm{l}+n)\big) \\ - \alpha \ln \pi_\theta\big(\boldsymbol{u}_\mathrm{l}(k_\mathrm{l}+n) \mid \boldsymbol{x}_\mathrm{rl}(k_\mathrm{l}+n)\big) \Big),
\end{aligned}
\end{equation*}
where $\psi$ is the index of the target critic, $\alpha$ is the entropy loss weight, and $\boldsymbol{u}_\mathrm{l} ( k_\mathrm{l}+n) \sim \pi_{\theta} \big(\cdot \mid \boldsymbol{x}_{\mathrm{rl}}(k_\mathrm{l}+n) \big)$.

The entropy loss weight is updated by minimizing the following loss function:
\begin{equation*}
    L(\alpha) = \frac{1}{N}\sum^{N}_{i=1} \big( -\alpha \ln \pi_\theta \big( \boldsymbol{u}_\mathrm{l}(i) \mid \boldsymbol{x}_\mathrm{rl}(i) \big) - \alpha \mathcal{H}
 \big),
\end{equation*}
where $\mathcal{H}<0$ is the target entropy used to adapt the entropy weight, with values closer to zero promoting a more stochastic policy \citep{haarnoja2018soft}.

The parameters $\phi_j$ of each critic, $j\in\{1,2\}$, are updated by minimizing the loss function in \eqref{eq:ddpg_loss}, with $\phi$ replaced by $\phi_j$ and $y$ given by the SAC target defined above, using the Adam optimizer. Then, the parameters of the actor, $\theta$, are subsequently updated using the policy gradient to maximize the entropy-regularized estimated return, defined as the sum of the minimum $Q$-value estimated by the two critics and the policy entropy weighted by $\alpha$ \citep{haarnoja2018soft}. The parameters of the target critics are updated using Polyak averaging as
\begin{equation*}
\begin{aligned}
\phi'_1 \leftarrow \tau \phi_1 + (1-\tau)\phi'_1, \\
\phi'_2 \leftarrow \tau \phi_2 + (1-\tau)\phi'_2.
\end{aligned}
\end{equation*}

The training procedure for the DRL-MPC framework is summarized in Algorithm~\ref{alg:training}.

\subsection{Network architectures and hyperparameters}
\label{subsec:hyperparams}

The DDPG actor and critic network architectures and hyperparameters are adopted from \citet{sun2024novel}, where they were shown to be effective for DDPG-based ramp metering control. The actor uses two hidden layers of width $256$ (ReLU activations), followed by a $\tanh$ output layer and an affine scaling to bound each action (i.e., each ramp metering rate) to $[0,1]$. The critic uses separate state and action paths with hidden layer widths $256$ and $128$, respectively. The concatenated features are passed through fully connected layers of widths $256$ and $128$ with ReLU activations, followed by a scalar output. The target actor and target critic use the same network architectures.

The SAC agent uses two critic networks with the same architecture as the DDPG critic, and the target critics use the same architecture. The SAC actor consists of a shared feature extractor with a hidden layer of width $256$ (ReLU activations), followed by separate mean and standard deviation heads, each with a hidden layer of width $256$. The standard deviation output uses a softplus activation to enforce nonnegative values. Sampled actions are mapped to the action bounds using the same $\tanh$ and affine scaling used for the DDPG actor. The hidden layer width of $256$ is chosen to match that of the DDPG actor.

Each agent is trained for $M=3500$ episodes. The DDPG and SAC agents use a replay buffer of length $2\cdot 10^{5}$, a mini-batch size $N=512$, discount factor $\gamma=0.99$, target smoothing factor $\tau = 10^{-2}$, and $n$-step TD targets with $n=10$, following \citet{sun2024novel}.

The actor and critic learning rates are set to $10^{-3}$ with gradient threshold $1$, and DDPG exploration uses an Ornstein-Uhlenbeck process with initial standard deviation $0.3$ and decay rate $5\cdot 10^{-6}$, as in \citet{sun2024novel}. For SAC, the entropy weight is initialized as $\alpha=1$ and is tuned during training using the target entropy $\mathcal{H}=-2$, set as the negative of the action dimension, similar to \citet{haarnoja2018soft2}. Learning is performed at the end of each episode once the replay buffer contains at least $N_\mathrm{w}=512$ samples. Each learning iteration comprises $N_\mathrm{e}=1$ epoch with at most $N_\mathrm{mb}^{\max}=100$ mini-batches, depending on the number available in the replay buffer, i.e., $N_\mathrm{mb}=\min\!\left\{N_\mathrm{mb}^{\max},\left\lfloor |\mathcal{D}|/N \right\rfloor\right\}$. Polyak updates with $\tau=10^{-2}$ are applied every $N_\mathrm{t}=10$ critic updates to the target actor and critic for DDPG and to the two target critics for SAC.

\section{Tuned PI-ALINEA parameters}\label{A2:PIalinea_params}
\begin{figure}[!htbp]
\centering
\includegraphics[width=0.9\textwidth]{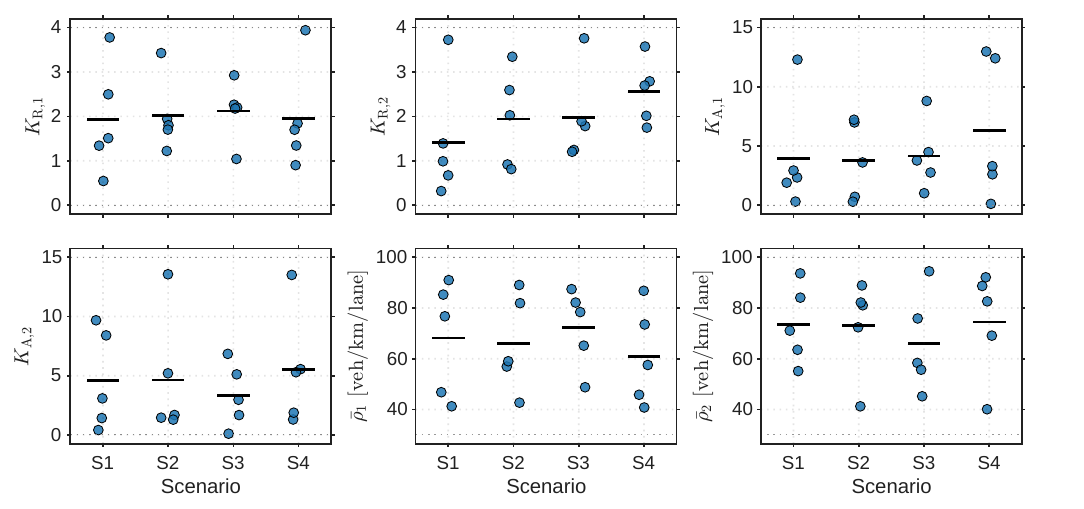}
\caption{Tuned PI-ALINEA parameters across the four scenarios. Each marker corresponds to one of the five independently tuned parameter sets per scenario, and the solid horizontal tick denotes the per-scenario mean.}
\label{fig:pi_alinea_params}
\end{figure}

Figure~\ref{fig:pi_alinea_params} reports the PI-ALINEA parameters obtained from the five independent Bayesian optimization runs per scenario described in Section~\ref{subsubsec:pi_alinea_mpc_case}.

\section*{Acknowledgements}
This research has received funding from the European Research Council (ERC) under the European Union's Horizon 2020 research and innovation programme (grant agreement No.~101018826, ERC Advanced Grant CLariNet).

\bibliography{ref.bib}

\end{document}